\documentclass[twocolumn,english,aps,prb,twocolumn,superscriptaddress,bibnotes,amsmath,amssymb,floatfix,groupedaddress,footinbib]{revtex4-2}
\usepackage[colorlinks=true,citecolor=blue,linkcolor=magenta]{hyperref}

\usepackage[flushleft]{threeparttable}
\usepackage[table]{xcolor}

\usepackage[utf8]{inputenc}
\usepackage[english]{babel}
\usepackage{amsmath,amsfonts,amssymb}
\usepackage[T1]{fontenc}
\usepackage{xurl}
\usepackage{soul}
\usepackage{changes}

\usepackage{amsmath}
\usepackage{amsfonts}
\usepackage{amssymb}
\usepackage[version=3]{mhchem}
\usepackage{stmaryrd}
\usepackage{epstopdf}
\usepackage{graphicx}
\graphicspath{{./Figures/}}

\begin{document}

\title{Single-shot Kramers–Kronig complex orbital angular momentum spectrum retrieval}

\author{Zhongzheng Lin$^{1,\ast}$, 
        Jianqi Hu$^{2,\ast,\dagger}$, 
        Yujie Chen$^1$, 
        Camille-Sophie Br\`es$^2$, 
        and Siyuan Yu$^{1,\ddag}$}
\affiliation{
$^1$State Key Laboratory of Optoelectronic Materials and Technologies, School of Electronics and Information Technology, Sun Yat-sen University, Guangzhou 510275, China.\\
$^2${\'E}cole Polytechnique F{\'e}d{\'e}rale de Lausanne, Photonic Systems Laboratory (PHOSL), STI-IEM, Station 11, Lausanne CH-1015, Switzerland.}

\maketitle

\noindent\textbf{\noindent
Orbital angular momentum (OAM) spectrum diagnosis is a fundamental building block for diverse OAM-based systems. Among others, the simple on-axis interferometric measurement can retrieve the amplitude and phase information of complex OAM spectra in a few shots. Yet, its single-shot retrieval remains illusive, due to the signal-signal beat interference inherent in the measurement. Here, we introduce the concept of Kramers-Kronig (KK) receiver in coherent communications to the OAM domain, enabling rigorous, single-shot OAM spectrum measurement. We explain in detail the working principle and the requirement of the KK method, and then apply the technique to precisely measure various characteristic OAM states. In addition, we discuss the effects of the carrier-to-signal power ratio and the number of sampling points essential for rigorous retrieval, and evaluate the performance on a large set of random OAM spectra and high-dimensional spaces. Single-shot KK interferometry shows enormous potential for characterizing complex OAM states in real-time. 
}

\section*{Introduction} 

\noindent{Structured} light waves with spiral phase fronts carry orbital angular momentum (OAM). Such helical optical beams, either naturally emitted from laser cavities \cite{naidoo2016controlled} or externally sculpted \cite{forbes2016creation}, have recently attracted significant attention and have been widely used in communication, sensing, imaging, as well as classical and quantum information processing \cite{willner2015optical,rubinsztein2016roadmap,padgett2017orbital,erhard2018twisted,shen2019optical}. 
For many of these applications, the ability to diagnose an arbitrary OAM state is essential, so that its complex structure can be unveiled and decomposed into orthogonal OAM basis. Various methods have been explored for the task. 
Perhaps the most straightforward approach is using vortex phase plates together with a mode filter to obtain the power of each OAM order sequentially \cite{forbes2016creation,schulze2013measurement}. The number of measurements required by this approach however scales fast with the measurement space. Techniques based on the far-field diffraction patterns through certain types of apertures \cite{hickmann2010unveiling} or gratings \cite{dai2015measuring,zheng2017measuring}, generally only identify pure OAM orders rather than superimposed states. 
In that sense, mode sorters are efficient in separating the superposition of OAM modes by mapping different OAM orders into different spatial positions. They have been implemented based on cascaded Mach-Zehnder interferometers with Dove prisms \cite{leach2002measuring}, log-polar transformation \cite{berkhout2010efficient,lavery2012refractive} and its improvement by means of beam-copying \cite{mirhosseini2013efficient}, or spiral transformation \cite{wen2018spiral} that increase the mode separation. 
Moreover, multi-plane light conversion is also employed for sorting OAM modes with enhanced functionalities \cite{labroille2014efficient,fontaine2019laguerre,zhang2020simultaneous}. 
Yet, mode sorters generally lose the relative phase information among OAM modes, which may be required in many scenarios to unambiguously reconstruct the complex OAM states. 
For this purpose the interference measurement techniques are appealing for the full OAM field retrieval \cite{huang2013phase,zhou2017orbital,d2017measuring,fu2020universal,kulkarni2017single}. 

Typical OAM interference measurements are performed by recording the interferograms of a complex signal field and a reference OAM mode (or Gaussian beam) with a camera \cite{zhou2017orbital,d2017measuring,fu2020universal}. 
Since the camera records the light intensity, the complex signal field cannot be directly retrieved from the intensity-only measurement. 
The measured interferogram intensity, however, consists of not only the desired interference between the signal and reference beams, but also the self-beating of both of them. 
While the power of the reference is constant across the azimuthal angle, the existence of the signal-signal beat interference (SSBI) complicates the signal retrieval process. 
Multiple interferograms are thus required to remove the SSBI, by adjusting the power \cite{zhou2017orbital,d2017measuring} and/or phase \cite{fu2020universal} of the reference light. Single-shot interferometric measurement has been demonstrated in the context of partially coherent fields, but only for symmetric OAM spectra, while two shots are still needed for generalized spectral shapes \cite{kulkarni2017single}.
% In addition, the characterization of complex OAM states has been realized through sequential weak and strong measurements \cite{malik2014direct}. 
Although the SSBI contribution may be omitted when the reference beam is sufficiently strong \cite{zhou2017orbital}, rigorous, single-shot retrieval remains elusive for conventional on-axis interferometry.

Notably, the interference measurement of the complex OAM spectrum resembles the detection of complex modulated signals in coherent optical communications, where the reference beam is the counterpart of the local oscillator. 
Phase-diverse coherent receiver is generally used to retrieve both the amplitude and phase of the modulated signal \cite{ip2008coherent}. In recent years, considerable efforts have been made to reduce the receiver complexity in coherent communication systems, ideally by using only one single-ended photodetector \cite{peng2009spectrally,randel2015100,mecozzi2016kramers,li2017ssbi,bo2018kramers,mecozzi2019kramers}. The Kramers-Kronig (KK) receiver is an effective solution for direct detection of complex-valued signals \cite{mecozzi2016kramers,li2017ssbi,bo2018kramers,mecozzi2019kramers}. 
In this case, the receiver works in a heterodyne scheme, and requires the frequency of the local oscillator being outside of the signal's spectrum. When the interfered waveform satisfies the minimum-phase condition \cite{mecozzi2019kramers}, the signal can be rigorously reconstructed from the intensity measurement via the famous KK relation. This greatly simplifies the receiver architecture into the straightforward direct detection. 
In addition, similar KK-based full-field retrieval has also been applied for holographic imaging exploring the space-time duality \cite{baek2019kramers}. 

In this work, by drawing a close analogy between the time-frequency and azimuth-OAM domains, we extend the KK retrieval concept into the OAM space, for the first time to our knowledge. Such an approach enables the readout of both the amplitude and phase relation of an arbitrary OAM state in a single-shot manner, without increasing the system complexity. We describe in detail the retrieval procedure, which is experimentally validated in a high-dimensional OAM space.
In particular, we demonstrate the diagnosis of a number of characteristic OAM states \cite{xie2017spatial,malik2014direct,hu2018talbot,lin2021spectral}, including fractional OAM modes \cite{gotte2008light} that slightly violates the KK retrieval criteria. 
The typical parameters essential for a KK receiver \cite{mecozzi2019kramers}, i.e., the carrier-to-signal power ratio (CSPR) and the number of sampling points, are discussed in the context of OAM fields. Furthermore, we compare the performances of the proposed KK approach and the conventional Fourier approach for a large set of arbitrary OAM states, where the superiority of the KK approach is clearly shown. 
The single-shot nature of the KK method may find useful applications for characterizing OAM-based systems in real-time.

%%%%%%%%%%%%%%%%%%%%%%%%%%%%%%%%%%%%%%%%%%%%
% FIGURE 1 
\begin{figure}[hbt!]
  \centering{
  \includegraphics{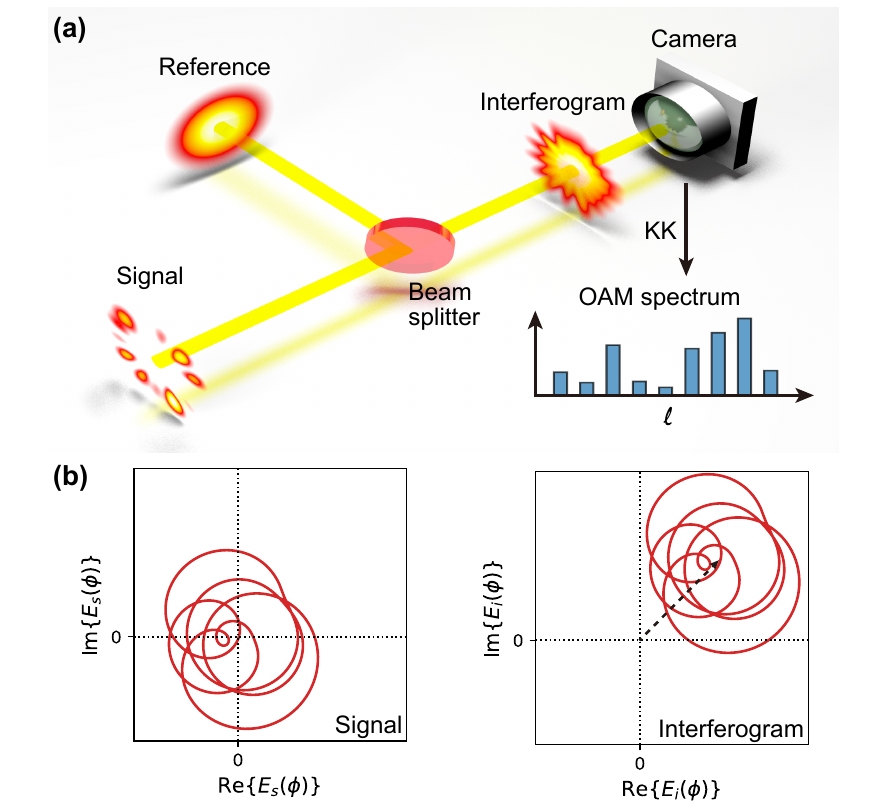}
  } 
    \caption{\noindent\textbf{Conceptual setup and the requirement for the Kramers-Kronig (KK) retrieval.} (a) The signal field with a complex orbital angular momentum (OAM) spectrum is co-axially combined with a reference Gaussian beam. The intensity of their interferogram is recorded with a camera and is used for the spectrum retrieval. (b) The azimuthal trajectories of the signal (left) and the interferogram (right) in the complex plane. In order to meet the minimum phase condition, the trajectory must not encircle the origin. With the addition of a sufficiently large reference field (denoted by the dashed arrow), the interferogram satisfies the requirement and thus the KK method can rigorously reconstruct the complex signal OAM spectrum. }
 \label{fig1}
\end{figure} 
%%%%%%%%%%%%%%%%%%%%%%%%%%%%%%%%%%%%%%%%%%%%%

\section*{Results} 
\subsection{Principle of operation} 
A complex OAM field can be described by the controlled superposition in discrete OAM mode basis \cite{xie2017spatial}:
\begin{equation}
E_s(\phi) = \sum_{l=l_0+1}^{l_0+N} a_l e^{il\phi + i\theta_l},
\label{eq:refname1}
\end{equation}
where $\phi$ is the azimuthal angle, $a_l$ and  $\theta_l$ are the amplitude and phase of the $l$-th OAM mode, respectively. 
Here we consider the modal decomposition of the OAM field in the interval of $[l_0+1, l_0+N]$, spanning over a $N$-dimensional space. Without loss of generality we assume $l_0 =0$ hereafter, as we can always use a phase mask with helical phase $e^{-il_0\phi}$ to shift the OAM spectrum entirely above the $0$-th order.  
In this study the OAM fields are constructed based on perfect vortex modes \cite{vaity2015perfect} with identical radial distributions (see Supplementary Note 1), such that we are only interested in field extraction in the azimuthal angle. 
As shown in Fig. \ref{fig1}(a), the complex signal field is interfered with a reference beam with plane wave front $E_r(\phi)=A e^{i\theta_r}$, where $A$ is the amplitude of the reference mode, and $\theta_r$ is the relative phase between the reference and signal fields. 
Similar to the single-sideband modulation for a KK receiver in communications \cite{mecozzi2016kramers}, no guard band is needed in between the reference and the signal OAM spectra. Then the interferogram is imaged with a camera and its intensity in the azimuthal angle is proportional to:
\begin{equation}
\begin{aligned}
|E_i(\phi)|^2 & = |E_s(\phi)+E_r(\phi)|^2 \\ 
& = A^2 + |E_s(\phi)|^2 + 2 A\operatorname{Re}\{E_s(\phi)e^{-i\theta_r}\} \\ 
& = A^2 + |\sum_{l=1}^{N} a_l e^{il\phi+i\theta_l}|^2 + 2 A \sum_{l=1}^{N} a_{l} \cos(l\phi + \theta_{l}-\theta_r).
\label{eq:refname2}
\end{aligned}
\end{equation}
The third term in the last equality of Eq. \eqref{eq:refname2} contains all the Fourier coefficients $a_l e^{i\theta_l}$ required to reconstruct the signal field, apart from scaling and a constant phase shift, while the second term corresponds to the SSBI. 

In order to extract the phase information from a single intensity measurement, the amplitude and phase of the interferogram $E_i(\phi)$ must be uniquely linked. A way to look at this is using the Z-extension as formulated in Ref. \cite{mecozzi2019kramers}. 
By replacing the $e^{i\phi}$ in $E_i(\phi)$ with a variable $Z$, the interferogram becomes a polynomial function:
\begin{equation}
\mathcal{Z}_{E_i}(Z)  = A e^{i\theta_r} + \sum_{l=1}^{N} a_l e^{i\theta_l}Z^{l}  =\frac{A e^{i\theta_r}}{\prod_{l=1}^{N} {(-z_l)}} \prod_{l=1}^{N} {(Z-z_l)},
\label{eq:refname3}
\end{equation}
where $z_l$ ($l=1,2,...,N$) are the roots of $\mathcal{Z}_{E_i}(Z)=0$, when $a_N \neq 0$. 
The second equality of Eq. \eqref{eq:refname3} shows that $\mathcal{Z}_{E_i}(Z)$ can be equivalently described by its zeros. 
Since the interferogram is under square-law detection, the zeros of $|E_i(\phi)|^2$ are found using properties of Z-extension \cite{mecozzi2019kramers}:
\begin{equation}
\mathcal{Z}_{|E_i|^2}(Z)  %\mathcal{Z}_{E_i}(Z)  \mathcal{Z}_{E_i^{*}}(Z)
=\frac{|A|^2}{Z^{N}\prod_{l=1}^{N} {(-z_l)}} \prod_{l=1}^{N} {(Z-z_l)(Z-\frac{1}{z_l^*})},
\label{eq:refname4}
\end{equation}
where $^{*}$ represents the complex conjugate. The zeros of $\mathcal{Z}_{|E_i|^2}(Z)$ are in $N$ pairs, comprised of the zeros of $\mathcal{Z}_{E_i}(Z)$ and the inverse of their complex conjugates. 
It is noted that replacing the zeros of $\mathcal{Z}_{E_i}(Z)$ with their inverse conjugates would not modify the function $\mathcal{Z}_{|E_i|^2}(Z)$. This implies multiple different interferograms are mapped to the same intensity profile and thus causing ambiguity for the retrieval. If the interferogram is constructed such that all its zeros are outside or on the unit circle, also known as the minimum phase waveform \cite{mecozzi2019kramers}, one-to-one mapping between $\mathcal{Z}_{E_i}(Z)$ and $\mathcal{Z}_{|E_i|^2}(Z)$ is established. 
Evidently, a necessary and sufficient condition to be of a minimum phase waveform is that its trajectory does not encircle the origin of the complex plane \cite{mecozzi2016necessary}. This is visualized in Fig. \ref{fig1}(b). The left panel illustrates the azimuthal trajectory of a general OAM signal field $E_s(\phi)$, which does not meet the minimum phase requirement. With a sufficiently large reference field $A e^{i\theta_r}$ (indicated by the dashed arrow), the azimuthal trajectory can be translated to match the minimum phase condition, as shown in the right panel of Fig. \ref{fig1}(b). Consequently, the amplitude and phase of the interferogram $E_i(\phi)$ is uniquely related. 
It is worth mentioning that, for a given signal OAM field, the required reference intensity for the minimum phase condition varies with the relative angle $\theta_r$. 
Nevertheless, since no prior knowledge of the signal field is known, the reference amplitude needs to be greater than the signal's peak amplitude in the azimuthal angle, i.e., $A>|E_s(\phi)|$ for $\phi \in [0,2\pi)$, thereby not encircling the origin. With this we can calculate the minimum CSPR required for rigorous retrieval, which is set by the peak to average power ratio of the signal field. 

Once the minimum phase condition is reached, for an interference field $E_i(\phi)=|E_i(\phi)|e^{i\arg(E_i(\phi))}$  ($\arg$ denotes the argument), the logarithm of its amplitude $\log(|E_i(\phi)|)$ and phase $\arg(E_i(\phi))$ is related by the Hilbert transform \cite{mecozzi2019kramers}: 
\begin{equation}
\arg(E_i(\phi)) = \frac{1}{2\pi}p.v.\int_0^{2\pi} \cot{\bigg(\frac{\phi-\phi^{\prime}}{2}\bigg)}\log(|E_i(\phi^{\prime})|) \,d\phi^{\prime},
\label{eq:refname5}
\end{equation}
where $p.v.$ is the principal value. Note that due to the periodic nature of the azimuthal space, the kernel here in the Hilbert transform is in the cotangent form rather than the inverse, and the integration is over one circle \cite{cizek1970discrete}. From Eq. \eqref{eq:refname5} we can reconstruct the full-field of the interferogram $E_i(\phi)$, and thus also the signal full-field $E_s(\phi)$ by removing the constant reference term. Based on the Fourier relation, this equivalently gives the amplitude and phase of the OAM spectrum of $E_s(\phi)$. 

\subsection{Experimental KK retrieval procedure} 
% %%%%%%%%%%%%%%%%%%%%%%%%%%%%%%%%%%%%%%%%%%%%
 \begin{figure*}[htp]
  \centering{
  \includegraphics{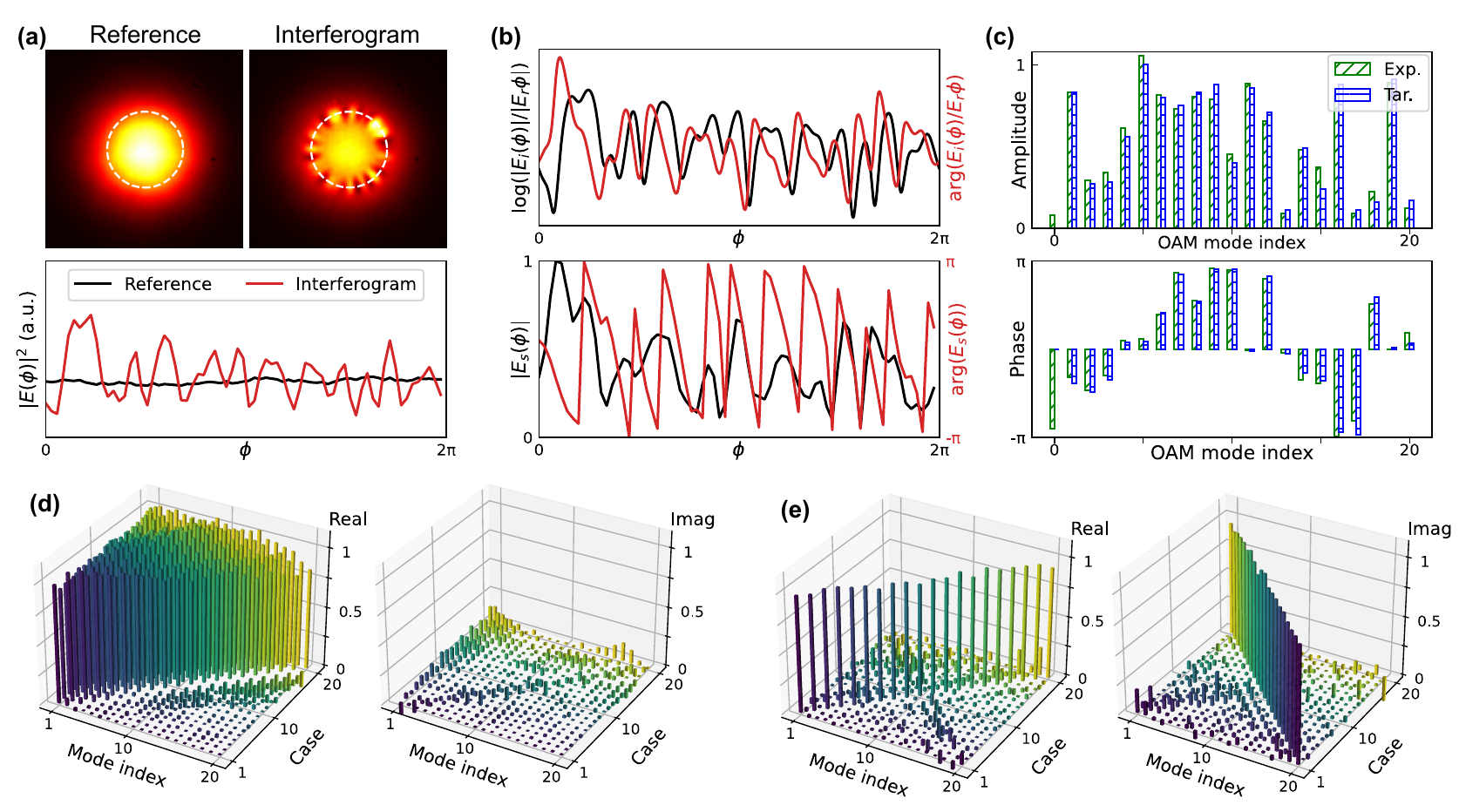}
  } \caption{\noindent\textbf{Experimental single-shot KK retrieval.} (a)-(c) Retrieval process of an arbitrary OAM spectrum. (a) The intensity images of the reference (top left) and the interferogram (top right) are captured by a camera. By sampling along the dashed circles in the images, their azimuthal distributions are extracted (bottom). (b) The interferogram is then normalized and digitally upsampled, whose phase is extracted from the Hilbert transform of the logarithm of its amplitude (top). From the full-field of the interferogram, the amplitude and phase of the signal is obtained and downsampled to the original sampling points (bottom). (c) Taking the Fourier transform of the signal field derives the normalized amplitude (top) and relative phase relation (bottom) of the OAM spectrum. An accuracy of $97.6\%$ is achieved by comparing the retrieved and the target OAM spectra. (d)-(e) Measured $2$D bar charts of the OAM states for (d) the superpositions of mode indices from $1$ to $n$ ($n= 1,2,...,20$) with equal amplitudes and in-phase relations; (e) the superpositions of mode indices $n$ and $21-n$ ($n= 1,2,...,20$) with equal amplitudes and a $\pi/2$ phase shift. The average retrieval accuracy in (d) and (e) are $98.9\%$ and $96.0\%$, respectively. 
 }
  \label{fig2}
\end{figure*} 
% %%%%%%%%%%%%%%%%%%%%%%%%%%%%%%%%%%%%%%%%%%%%
In the experiment, the complex signal OAM fields are synthesized using computer generated holograms and an optical 4-f system \cite{arrizon2007pixelated}. They consist in arbitrary superpositions of ring-shaped OAM modes with topological charges spanning from $1$ to $20$. The OAM fields under test are then co-axially combined with a reference Gaussian beam, and their interferograms are recorded with a camera. The detailed experimental setup is described in the Supplementary Note 1. To satisfy the minimum phase condition, we tune the reference light power slightly above the power threshold set by the minimum required CSPR. Thanks to the signal preparation method we used \cite{arrizon2007pixelated}, the experimental and minimum required CSPR difference is kept constant ($\sim 1~{\rm dB}$) for different measurement instances (see Supplementary Note 2). 
In the following, we demonstrate the KK retrieval procedure on a random complex OAM spectrum as an example. The top panel of Fig. \ref{fig2}(a) shows the measured camera images of the reference and the interference beams. Their azimuthal intensity distributions are extracted by sampling around the interfered regions (indicated by the dashed circles), and unwrapped as shown in the bottom panel of Fig. \ref{fig2}(a). It can be seen that the measured reference $|E_{r}(\phi)|^2$ shows small intensity fluctuations in the azimuth direction.
Such a reference intensity pattern is unchanged while characterizing different signal OAM fields, so only their interferogram $|E_{i}(\phi)|^2$ needs to be measured each time, suggesting that the retrieval process is a single-shot.

Similar to the KK full-field retrieval in space and time \cite{mecozzi2016kramers,baek2019kramers}, digital upsampling may also be required in our case, if the number of physical sampling points is not sufficiently large to cover the spectrum expended by the logarithmic operation \cite{mecozzi2019kramers}. The effect of upsampling is discussed in Supplementary Note 3. Throughout the paper we take $71$ azimuthal sampling points in the experiment, and then we digitally upsample the normalized interferogram $|E_{i}(\phi)|^2/|E_{r}(\phi)|^2$ by $11$-times for the subsequent processing. 
The top panel of the Fig. \ref{fig2}(b) shows the logarithm of the upsampled interferogram $\log(|E_{i}(\phi)|/|E_{r}(\phi)|)$ as well as its Hilbert transform $\arg(E_{i}(\phi)/E_{r}(\phi))$. Instead of taking the convolution as defined in Eq. \eqref{eq:refname5}, the actual implementation of the Hilbert transform is realized in the spectral domain using the fast Fourier transform and the sign function \cite{mecozzi2016kramers}. 
The full-field of the interferogram $E_{i}(\phi)/E_{r}(\phi)$ is thus derived, from which the full signal field can be easily calculated by $E_{s}(\phi) \approx |E_{r}(\phi)|(E_{i}(\phi)/E_{r}(\phi)-1)$. Here we omit the negligible phase non-uniformity of the experimental reference beam. The retrieved signal field is then downsampled, whose normalized amplitude $|E_{s}(\phi)|$ and phase $\arg(E_{s}(\phi))$ are plotted in the bottom panel of Fig. \ref{fig2}(b). Finally, the complex signal OAM spectrum is retrieved by taking the Fourier transform of $E_{s}(\phi)$. Figure \ref{fig2}(c) illustrates the retrieved amplitude (top) and phase (bottom) profile of the OAM spectrum and are compared to the ground truth. 
It can be seen that they are in excellent agreement. To quantitatively assess the performance of the KK retrieval, we introduce the following metric based on the overlap integral of the target and the retrieved OAM spectra: 
\begin{equation}
 \eta = \frac{|\sum_{-\infty}^{\infty} a_l a_l^{\prime} e^{i\theta_l-i\theta_l^{\prime}}|^2}{(\sum_{-\infty}^{\infty} |a_l|^2)(\sum_{-\infty}^{\infty} |a_l^{\prime}|^2)},
\label{eq:refname6}
\end{equation}
where $a_l^{\prime}$ and $\theta_l^{\prime}$ are the amplitude and phase of the $l$-th order OAM mode of the retrieved OAM spectrum, respectively. In Fig. \ref{fig2}(c), the retrieval accuracy $\eta $ is found to be $97.6\%$.

To further test the validity of the proposed method, we apply the KK retrieval to diagnose a series of OAM spectra shown in Figs. \ref{fig2}(d) and \ref{fig2}(e). The first scenario is the rectangular OAM spectra $E_s(\phi) = \sum_{l=1}^{n}e^{il\phi}$ of different widths ($n= 1,2,...,20$), where the constituent OAM modes are of equal amplitudes and in-phase relations. Figure \ref{fig2}(d) shows the real and imaginary parts of the measured OAM spectra for these cases in a complex $2$D bar chart. As expected, the retrieved OAM spectra are approximately of rectangular shapes, and are mainly populated in the real part of the bar chart due to their in-phase features. An average accuracy of $98.9\%$ is achieved for these measurements. 
For the second scenario, the OAM spectra are structured by $E_s(\phi) = e^{in\phi} + i e^{i(21-n)\phi}$, where $n= 1,2,...,20$. In this case, the bar chart should be diagonal in its real part while anti-diagonal in its imaginary part. This is clearly observed in the measurements shown in Fig. \ref{fig2}(e), with an average retrieval accuracy reaching $96.0\%$. 

\begin{figure*}[ht!]
\centering\includegraphics{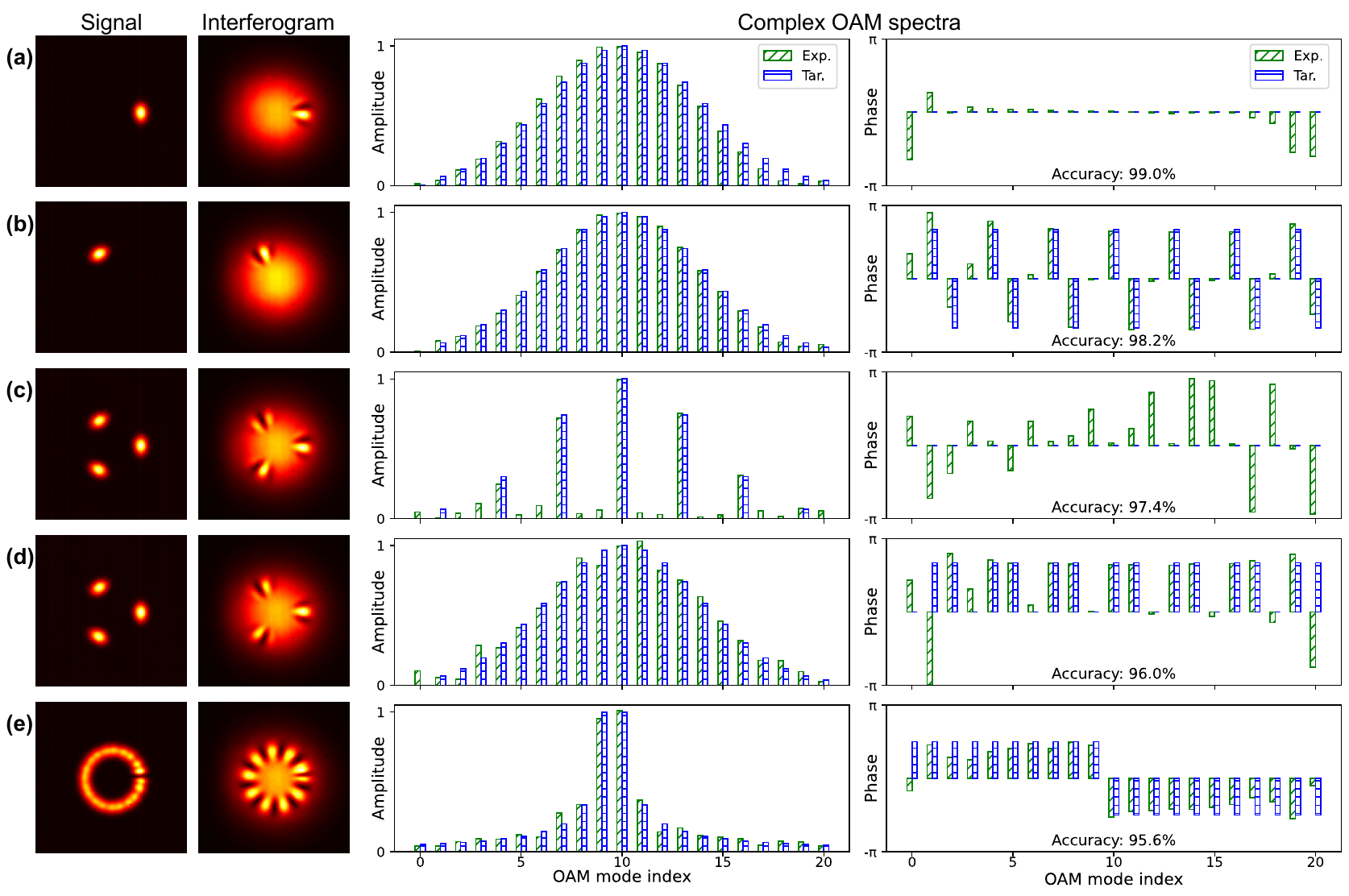}
\caption{\noindent\textbf{Measurements of complex OAM spectra.} Left: the measured intensity images of the signal and the interferogram; Right: the normalized amplitude and relative phase of the target and experimentally retrieved OAM spectrum. (a)-(d) Gaussian-shaped OAM spectra centered at $10$-th order with versatile OAM mode spacings and relative phase relations. (a) In-phase OAM spectrum with a mode spacing of $1$. (b) Linear-phase OAM spectrum with a phase slope of $2\pi/3$ and a mode spacing of $1$. (c) In-phase OAM spectrum with a mode spacing of $3$. (d) OAM spectrum with periodic Talbot phase $[0, 2\pi/3, 2\pi/3]$ and a mode spacing of $1$. (e) Fractional OAM mode with a topological charge of $9.5$. The retrieval accuracy is also indicated for each case. }
\label{fig3}
\end{figure*}

\subsection{Measurement of characteristic OAM states}
The single-shot KK full-field retrieval allows for the simple characterization of complex OAM states. In the following, we study various characteristic OAM spectra displayed in Fig. \ref{fig3}. Figure \ref{fig3}(a) demonstrates the measurement of a signal field with a Gaussian-shaped, in-phase OAM spectrum centered at $10$-th order, i.e., $E_s(\phi) = \sum_{l=1}^{20}e^{-\frac{(l-10)^2}{30}}e^{il\phi}$, which corresponds to a bright petal horizontally aligned in the azimuthal angle \cite{xie2017spatial}. From only a single interferogram we can reconstruct the signal's OAM spectrum well matching the ground truth. Note that all the measured images of the signal beams in Fig. \ref{fig3} are displayed only for the illustration purpose, and are not used for their OAM spectra retrieval. Figure \ref{fig3}(b) shows the condition for the same petal field as in Fig. \ref{fig3}(a) except being rotated by $2\pi/3$ counterclockwise. 
A linear phase ramp is thus imparted to the OAM spectrum due to the rotation, with a phase slope of $2\pi/3$. This feature is well captured by the KK approach in a single-shot fashion, simplifying the previously used schemes like the sequential weak and strong measurements \cite{malik2014direct} or the phase-shifting holograms \cite{xie2017using}. 

In addition, it is known that increasing the mode spacing in the OAM spectrum will lead to the multiplication of petals in the azimuthal angle \cite{xie2017spatial}. This is shown in Fig. \ref{fig3}(c), where a $3$-petal field under test is constructed from in-phase OAM modes with an order spacing of $3$ and the same envelope as in Figs. \ref{fig3}(a) and \ref{fig3}(b). The interferogram could accurately retrieve its equidistant OAM structure as well as its in-phase relation. Notice that the large phase error in Fig. \ref{fig3}(c) is only associated with void OAM modes or at small modal amplitudes. Furthermore, we investigate the complex OAM spectrum of the petal field in Fig. \ref{fig3}(c) being subjected to phase modulation. Specifically, when the $3$ petals are modulated with the Talbot phase sequence of $[0, -2\pi/3,-2\pi/3]$, the initial OAM spectrum is self-imaged to create new OAM modes, meanwhile preserving its overall envelope \cite{lin2021spectral}. The measurement results are shown in Fig. \ref{fig3}(d). Although the signal intensity patterns in Figs. \ref{fig3}(c) and \ref{fig3}(d) are identical, the phenomenon of the OAM spectral self-imaging is clearly observed from the KK retrieval. More interestingly, the approach also provides a direct phase measurement of all the OAM modes. It can be seen that the relative phases of OAM self-images again follow the Talbot relation of $[0, 2\pi/3, 2\pi/3]$, apart from a constant phase (subtracted thus not shown in Fig. \ref{fig3}(d) for better representability). Notably, such a phase structure of Talbot self-images has only been recently determined in space \cite{de2015phases} and time \cite{clement2020far}, while here we measure it for the first time in the OAM basis. 

We also use the KK method to characterize the complex spectra of fractional OAM modes. A fractional OAM order can be viewed as the weighted superposition of integer OAM modes \cite{zhang2022review}. In Fig. \ref{fig3}(e), we show the measurement of the fractional OAM field with a topological charge of $9.5$, i.e., $E_s(\phi) = e^{i9.5\phi}$. 
The field intrinsically exhibits around $1\%$ power leakage to the negative OAM orders, thus not rigorously satisfying the single-sideband requirement. 
Nevertheless, since the leakage is small, the KK retrieval still works effectively with the accuracy of $95.6\%$. We can see that the fractional OAM field in Fig. \ref{fig3}(e) is mainly composed of the $9$-th and $10$-th OAM modes, together with the other OAM orders slowly decaying when moving away from the center modes. Moreover, we are able to identify the phase relation of its constituent OAM modes, which is rarely explored for the fractional OAM states. It is found that the orders at left and right parts of the fractional OAM spectrum are approximately out-of-phase. The phase deviation between the measured and theoretical fractional OAM spectra may come in part from the finite sampling in the azimuthal angle. 

\subsection{Effect of CSPR levels}
\begin{figure}[ht!]
\centering\includegraphics{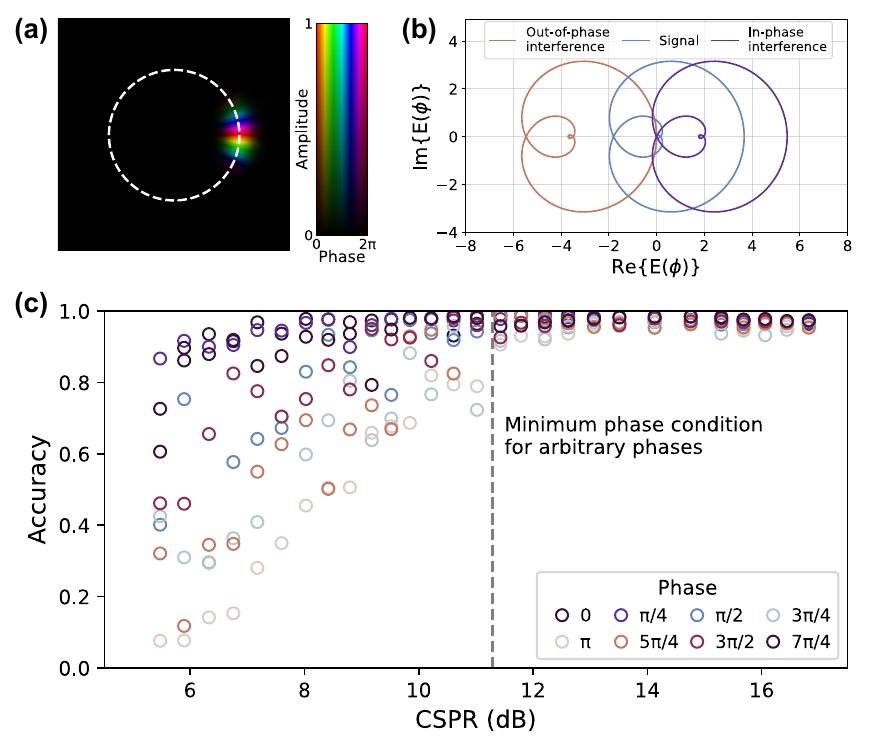}
\caption{\noindent\textbf{KK retrieval at different carrier-to-signal power ratio (CSPR) levels.} (a) The complex amplitude field of the signal used for the study [same as the signal field in Fig. \ref{fig3}(a)]. 
(b) The azimuthal trajectories of the signal field and interferograms with in-phase and out-of-phase addition of the reference field. The trajectories of interferograms are exactly at the limit of the minimum phase condition, which correspond to the minimum required CSPRs of $5.1~{\rm dB}$ (in-phase) and $11.3~{\rm dB}$ (out-of-phase), respectively. (c) The accuracy of the KK retrieval at different CSPR levels and varying phases between the signal and reference fields. Below the CSPR threshold for arbitrary phases ($11.3~{\rm dB}$, indicated by the dashed line), the retrieval performance varies with the phase; while above, the retrieval accuracy is approximately close to the unity for all the phases. 
}
\label{fig4}
\end{figure}

Figure \ref{fig4} shows the KK retrieval performance at different CSPR levels. As an example, the same OAM state in Fig. \ref{fig3}(a) is used as the signal field for the study, whose complex amplitude field is displayed in Fig. \ref{fig4}(a). From this we can plot in Fig. \ref{fig4}(b) the trajectory of its azimuthal distribution in the complex plane. As mentioned previously, to guarantee the rigorous KK retrieval, the trajectory of the signal field after interfering with the reference light must not encircle the origin of the complex plane. The minimum required reference intensity for such a criterion is highly dependant on the relative angle between the reference and the signal fields. 
In Fig. \ref{fig4}(b), the minimum required CSPR varies in a range between $5.1~{\rm dB}$ and $11.3~{\rm dB}$, where the lower and upper bounds correspond to the reference field being added in-phase ($\theta_r = 0$) and out-of-phase ($\theta_r = \pi$) with the signal field, respectively. The minimum required CSPR value valid for all these relative angles is thus set by the upper bound, i.e., $11.3~{\rm dB}$. 

Figure \ref{fig4}(c) presents the experimental results for the retrieval accuracy at different CSPR levels and various relative phases between the signal and reference fields. In the experiment, the control of the CSPR and their relative phase are realized via the computer generated holograms. The phases shown in Fig. \ref{fig4}(c) are varied in steps of $\pi/4$. We emphasize here that they do not correspond to the actual $\theta_r$, but are offset from a constant unknown phase due to experimental constraints. In Fig. \ref{fig4}(c), 
when the experimental CSPR is well below the $11.3~{\rm dB}$ threshold (marked by the dashed line), the retrieval performance changes significantly with the phase, although for some angles the retrieval accuracy are acceptable. Once the CSPR exceeds the threshold, decent retrieval is achieved for arbitrary relative phases between the reference and signal fields. The experimental results in Fig. \ref{fig4}(c) are in accordance with the theoretical analysis carried out above. 

\subsection{Performance evaluation}
\begin{figure}[ht!]
\centering\includegraphics{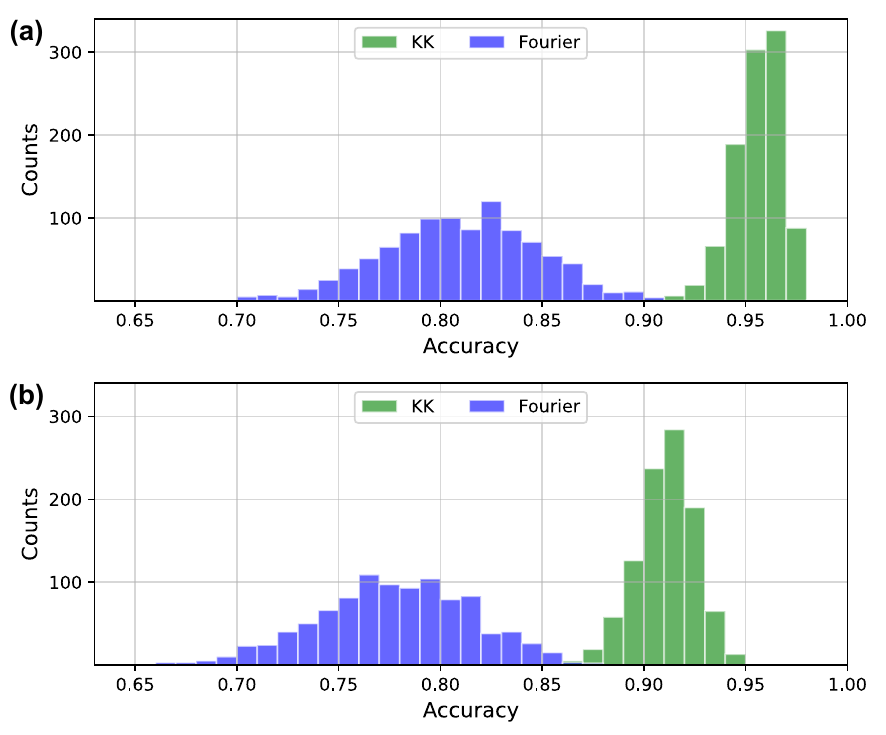}
\caption{\noindent\textbf{Performance evaluation of the KK retrieval on random OAM spectra.} (a)-(b) Histograms of the retrieval accuracy of the KK method and the conventional Fourier method, measured on $1000$ OAM spectra with random complex mode coefficients. (a) For an OAM measurement range from $1$ to $20$, the average and standard deviation of the KK retrieval accuracy are $95.6\%$ and $1.2\%$, respectively. (b) For an OAM measurement range from $1$ to 30, the average and standard deviation of the KK retrieval accuracy are $91.1\%$ and $1.4\%$, respectively. The KK method shows superiority over the Fourier method in both cases. 
}
\label{fig5}
\end{figure}
In this part, we evaluate the performance of the KK retrieval on a large set of OAM spectra generated with random complex mode coefficients. 
As in the previous measurements, the difference between the experimental and minimum required CSPRs is automatically maintained around $1~{\rm dB}$, which is experimentally confirmed in Supplementary Note 2 for $100$ random OAM spectra. Figure \ref{fig5}(a) shows the histogram of the KK retrieval accuracy for $1000$ spectrum samples on the same dimensional space as before. An average retrieval accuracy of $95.6\%$ is obtained with a standard deviation of $1.2\%$. The performance of the KK retrieval is also compared with the conventional Fourier method, computed by the Fourier transform disregarding the SSBI in Eq. \eqref{eq:refname2}. A clear advantage of using the KK method can be seen in Fig. \ref{fig5}(a). Next, we further push the measurement dimensionality up to $30$-th OAM order, while keeping the azimuthal sampling points and the digital upsampling unchanged. Figure \ref{fig5}(b) shows the corresponding experimental results. The average KK retrieval accuracy  in this case still reaches $91.1\%$ with a standard deviation of $1.4\%$, outperforming the conventional Fourier method by a large margin. Although the performance of the Fourier method may be improved by increasing the reference power, keeping relatively low CSPR values is favored to avoid large DC components in detection and thus maximally utilize the dynamic range of the camera. 
\vspace{0.5cm}
\section*{Discussion} 
The experimental setup used in this work is a conventional on-axis interferometer equivalent to the configurations in Refs.  \cite{zhou2017orbital,d2017measuring,fu2020universal}. 
However, contrary to all the past demonstrations that require a few shots to diagnose a complex OAM spectrum, our method provides single-shot retrieval mediated by the famous KK relation. This greatly accelerate the measurement as it bypasses the need to adjust the amplitude and/or phase of the reference when characterizing each superimposed state \cite{zhou2017orbital,d2017measuring,fu2020universal}. In our system, the speed of the measurement is defined by the frame rate of the camera. Since in this study we are dealing with only the azimuthal field distribution, the detection can be seamlessly connected to the rotational Doppler effect \cite{courtial1998measurement}. In this scenario, the camera is replaced by a fast photodetector with a spinning phase mask performing the azimuth-to-time mapping \cite{zhou2017orbital}. 

To sum up this work, we propose and experimentally demonstrate a high-dimensional OAM analyzer that can measure complex OAM states in one single shot. The idea is inspired by the KK receiver in optical communications, while here we introduce the same concept to the OAM spectrum analysis. As demonstrated here, this enables the simple characterization of a wide variety of complex OAM spectra, and can be extended to the measurements of multiple concentric OAM states as well as, in principle, the superposition of Laguerre-Gaussian modes. 
The proposed single-shot KK interferometry can be readily employed for state measurements in OAM-based information processing, sensing, and communication systems. 
This work also implies the general duality between azimuth-OAM and time-frequency, suggesting that their processing techniques may be borrowed interchangeably.

\medskip
\begin{footnotesize}

\noindent \textbf{Funding}: National Key Research and Development Program of China (2018YFB1801803, 2019YFA0706302); Basic and Applied Basic Research Foundation of Guangdong Province (2021B1515020093, 2021B1515120057); Local Innovative and Research Teams Project of Guangdong Pearl River Talents Program (2017BT01X121); Swiss National Science Foundation (P2ELP2$\_$199825).

\vspace{0.1cm}

\noindent \textbf{Data Availability Statement}: 
The data and code that support the plots within this paper and other findings of this study are available from the corresponding authors upon reasonable request.
\end{footnotesize}

\renewcommand{\bibpreamble}{
$^\ast$These authors contributed equally to this work.\\
$^\dagger${Corresponding author: \textcolor{magenta}{jianqi.hu@epfl.ch}}\\
$^\ddag${Email: \textcolor{magenta}{yusy@mail.sysu.edu.cn}}
}

 \bibliographystyle{apsrev4-2}
\bibliography{ref}

\begin{thebibliography}{46}%
\makeatletter
\providecommand \@ifxundefined [1]{%
 \@ifx{#1\undefined}
}%
\providecommand \@ifnum [1]{%
 \ifnum #1\expandafter \@firstoftwo
 \else \expandafter \@secondoftwo
 \fi
}%
\providecommand \@ifx [1]{%
 \ifx #1\expandafter \@firstoftwo
 \else \expandafter \@secondoftwo
 \fi
}%
\providecommand \natexlab [1]{#1}%
\providecommand \enquote  [1]{``#1''}%
\providecommand \bibnamefont  [1]{#1}%
\providecommand \bibfnamefont [1]{#1}%
\providecommand \citenamefont [1]{#1}%
\providecommand \href@noop [0]{\@secondoftwo}%
\providecommand \href [0]{\begingroup \@sanitize@url \@href}%
\providecommand \@href[1]{\@@startlink{#1}\@@href}%
\providecommand \@@href[1]{\endgroup#1\@@endlink}%
\providecommand \@sanitize@url [0]{\catcode `\\12\catcode `\$12\catcode
  `\&12\catcode `\#12\catcode `\^12\catcode `\_12\catcode `\%12\relax}%
\providecommand \@@startlink[1]{}%
\providecommand \@@endlink[0]{}%
\providecommand \url  [0]{\begingroup\@sanitize@url \@url }%
\providecommand \@url [1]{\endgroup\@href {#1}{\urlprefix }}%
\providecommand \urlprefix  [0]{URL }%
\providecommand \Eprint [0]{\href }%
\providecommand \doibase [0]{https://doi.org/}%
\providecommand \selectlanguage [0]{\@gobble}%
\providecommand \bibinfo  [0]{\@secondoftwo}%
\providecommand \bibfield  [0]{\@secondoftwo}%
\providecommand \translation [1]{[#1]}%
\providecommand \BibitemOpen [0]{}%
\providecommand \bibitemStop [0]{}%
\providecommand \bibitemNoStop [0]{.\EOS\space}%
\providecommand \EOS [0]{\spacefactor3000\relax}%
\providecommand \BibitemShut  [1]{\csname bibitem#1\endcsname}%
\let\auto@bib@innerbib\@empty
%</preamble>
\bibitem [{\citenamefont {Naidoo}\ \emph {et~al.}(2016)\citenamefont {Naidoo},
  \citenamefont {Roux}, \citenamefont {Dudley}, \citenamefont {Litvin},
  \citenamefont {Piccirillo}, \citenamefont {Marrucci},\ and\ \citenamefont
  {Forbes}}]{naidoo2016controlled}%
  \BibitemOpen
  \bibfield  {author} {\bibinfo {author} {\bibfnamefont {D.}~\bibnamefont
  {Naidoo}}, \bibinfo {author} {\bibfnamefont {F.~S.}\ \bibnamefont {Roux}},
  \bibinfo {author} {\bibfnamefont {A.}~\bibnamefont {Dudley}}, \bibinfo
  {author} {\bibfnamefont {I.}~\bibnamefont {Litvin}}, \bibinfo {author}
  {\bibfnamefont {B.}~\bibnamefont {Piccirillo}}, \bibinfo {author}
  {\bibfnamefont {L.}~\bibnamefont {Marrucci}},\ and\ \bibinfo {author}
  {\bibfnamefont {A.}~\bibnamefont {Forbes}},\ }\href@noop {} {\bibfield
  {journal} {\bibinfo  {journal} {Nature Photonics}\ }\textbf {\bibinfo
  {volume} {10}},\ \bibinfo {pages} {327} (\bibinfo {year} {2016})}\BibitemShut
  {NoStop}%
\bibitem [{\citenamefont {Forbes}\ \emph {et~al.}(2016)\citenamefont {Forbes},
  \citenamefont {Dudley},\ and\ \citenamefont {McLaren}}]{forbes2016creation}%
  \BibitemOpen
  \bibfield  {author} {\bibinfo {author} {\bibfnamefont {A.}~\bibnamefont
  {Forbes}}, \bibinfo {author} {\bibfnamefont {A.}~\bibnamefont {Dudley}},\
  and\ \bibinfo {author} {\bibfnamefont {M.}~\bibnamefont {McLaren}},\
  }\href@noop {} {\bibfield  {journal} {\bibinfo  {journal} {Advances in Optics
  and Photonics}\ }\textbf {\bibinfo {volume} {8}},\ \bibinfo {pages} {200}
  (\bibinfo {year} {2016})}\BibitemShut {NoStop}%
\bibitem [{\citenamefont {Willner}\ \emph {et~al.}(2015)\citenamefont
  {Willner}, \citenamefont {Huang}, \citenamefont {Yan}, \citenamefont {Ren},
  \citenamefont {Ahmed}, \citenamefont {Xie}, \citenamefont {Bao},
  \citenamefont {Li}, \citenamefont {Cao}, \citenamefont {Zhao} \emph
  {et~al.}}]{willner2015optical}%
  \BibitemOpen
  \bibfield  {author} {\bibinfo {author} {\bibfnamefont {A.~E.}\ \bibnamefont
  {Willner}}, \bibinfo {author} {\bibfnamefont {H.}~\bibnamefont {Huang}},
  \bibinfo {author} {\bibfnamefont {Y.}~\bibnamefont {Yan}}, \bibinfo {author}
  {\bibfnamefont {Y.}~\bibnamefont {Ren}}, \bibinfo {author} {\bibfnamefont
  {N.}~\bibnamefont {Ahmed}}, \bibinfo {author} {\bibfnamefont
  {G.}~\bibnamefont {Xie}}, \bibinfo {author} {\bibfnamefont {C.}~\bibnamefont
  {Bao}}, \bibinfo {author} {\bibfnamefont {L.}~\bibnamefont {Li}}, \bibinfo
  {author} {\bibfnamefont {Y.}~\bibnamefont {Cao}}, \bibinfo {author}
  {\bibfnamefont {Z.}~\bibnamefont {Zhao}}, \emph {et~al.},\ }\href@noop {}
  {\bibfield  {journal} {\bibinfo  {journal} {Advances in Optics and
  Photonics}\ }\textbf {\bibinfo {volume} {7}},\ \bibinfo {pages} {66}
  (\bibinfo {year} {2015})}\BibitemShut {NoStop}%
\bibitem [{\citenamefont {Rubinsztein-Dunlop}\ \emph
  {et~al.}(2016)\citenamefont {Rubinsztein-Dunlop}, \citenamefont {Forbes},
  \citenamefont {Berry}, \citenamefont {Dennis}, \citenamefont {Andrews},
  \citenamefont {Mansuripur}, \citenamefont {Denz}, \citenamefont {Alpmann},
  \citenamefont {Banzer}, \citenamefont {Bauer} \emph
  {et~al.}}]{rubinsztein2016roadmap}%
  \BibitemOpen
  \bibfield  {author} {\bibinfo {author} {\bibfnamefont {H.}~\bibnamefont
  {Rubinsztein-Dunlop}}, \bibinfo {author} {\bibfnamefont {A.}~\bibnamefont
  {Forbes}}, \bibinfo {author} {\bibfnamefont {M.~V.}\ \bibnamefont {Berry}},
  \bibinfo {author} {\bibfnamefont {M.~R.}\ \bibnamefont {Dennis}}, \bibinfo
  {author} {\bibfnamefont {D.~L.}\ \bibnamefont {Andrews}}, \bibinfo {author}
  {\bibfnamefont {M.}~\bibnamefont {Mansuripur}}, \bibinfo {author}
  {\bibfnamefont {C.}~\bibnamefont {Denz}}, \bibinfo {author} {\bibfnamefont
  {C.}~\bibnamefont {Alpmann}}, \bibinfo {author} {\bibfnamefont
  {P.}~\bibnamefont {Banzer}}, \bibinfo {author} {\bibfnamefont
  {T.}~\bibnamefont {Bauer}}, \emph {et~al.},\ }\href@noop {} {\bibfield
  {journal} {\bibinfo  {journal} {Journal of Optics}\ }\textbf {\bibinfo
  {volume} {19}},\ \bibinfo {pages} {013001} (\bibinfo {year}
  {2016})}\BibitemShut {NoStop}%
\bibitem [{\citenamefont {Padgett}(2017)}]{padgett2017orbital}%
  \BibitemOpen
  \bibfield  {author} {\bibinfo {author} {\bibfnamefont {M.~J.}\ \bibnamefont
  {Padgett}},\ }\href@noop {} {\bibfield  {journal} {\bibinfo  {journal}
  {Optics Express}\ }\textbf {\bibinfo {volume} {25}},\ \bibinfo {pages}
  {11265} (\bibinfo {year} {2017})}\BibitemShut {NoStop}%
\bibitem [{\citenamefont {Erhard}\ \emph {et~al.}(2018)\citenamefont {Erhard},
  \citenamefont {Fickler}, \citenamefont {Krenn},\ and\ \citenamefont
  {Zeilinger}}]{erhard2018twisted}%
  \BibitemOpen
  \bibfield  {author} {\bibinfo {author} {\bibfnamefont {M.}~\bibnamefont
  {Erhard}}, \bibinfo {author} {\bibfnamefont {R.}~\bibnamefont {Fickler}},
  \bibinfo {author} {\bibfnamefont {M.}~\bibnamefont {Krenn}},\ and\ \bibinfo
  {author} {\bibfnamefont {A.}~\bibnamefont {Zeilinger}},\ }\href@noop {}
  {\bibfield  {journal} {\bibinfo  {journal} {Light: Science \& Applications}\
  }\textbf {\bibinfo {volume} {7}},\ \bibinfo {pages} {17146} (\bibinfo {year}
  {2018})}\BibitemShut {NoStop}%
\bibitem [{\citenamefont {Shen}\ \emph {et~al.}(2019)\citenamefont {Shen},
  \citenamefont {Wang}, \citenamefont {Xie}, \citenamefont {Min}, \citenamefont
  {Fu}, \citenamefont {Liu}, \citenamefont {Gong},\ and\ \citenamefont
  {Yuan}}]{shen2019optical}%
  \BibitemOpen
  \bibfield  {author} {\bibinfo {author} {\bibfnamefont {Y.}~\bibnamefont
  {Shen}}, \bibinfo {author} {\bibfnamefont {X.}~\bibnamefont {Wang}}, \bibinfo
  {author} {\bibfnamefont {Z.}~\bibnamefont {Xie}}, \bibinfo {author}
  {\bibfnamefont {C.}~\bibnamefont {Min}}, \bibinfo {author} {\bibfnamefont
  {X.}~\bibnamefont {Fu}}, \bibinfo {author} {\bibfnamefont {Q.}~\bibnamefont
  {Liu}}, \bibinfo {author} {\bibfnamefont {M.}~\bibnamefont {Gong}},\ and\
  \bibinfo {author} {\bibfnamefont {X.}~\bibnamefont {Yuan}},\ }\href@noop {}
  {\bibfield  {journal} {\bibinfo  {journal} {Light: Science \& Applications}\
  }\textbf {\bibinfo {volume} {8}},\ \bibinfo {pages} {1} (\bibinfo {year}
  {2019})}\BibitemShut {NoStop}%
\bibitem [{\citenamefont {Schulze}\ \emph {et~al.}(2013)\citenamefont
  {Schulze}, \citenamefont {Dudley}, \citenamefont {Flamm}, \citenamefont
  {Duparre},\ and\ \citenamefont {Forbes}}]{schulze2013measurement}%
  \BibitemOpen
  \bibfield  {author} {\bibinfo {author} {\bibfnamefont {C.}~\bibnamefont
  {Schulze}}, \bibinfo {author} {\bibfnamefont {A.}~\bibnamefont {Dudley}},
  \bibinfo {author} {\bibfnamefont {D.}~\bibnamefont {Flamm}}, \bibinfo
  {author} {\bibfnamefont {M.}~\bibnamefont {Duparre}},\ and\ \bibinfo {author}
  {\bibfnamefont {A.}~\bibnamefont {Forbes}},\ }\href@noop {} {\bibfield
  {journal} {\bibinfo  {journal} {New Journal of Physics}\ }\textbf {\bibinfo
  {volume} {15}},\ \bibinfo {pages} {073025} (\bibinfo {year}
  {2013})}\BibitemShut {NoStop}%
\bibitem [{\citenamefont {Hickmann}\ \emph {et~al.}(2010)\citenamefont
  {Hickmann}, \citenamefont {Fonseca}, \citenamefont {Soares},\ and\
  \citenamefont {Ch{\'a}vez-Cerda}}]{hickmann2010unveiling}%
  \BibitemOpen
  \bibfield  {author} {\bibinfo {author} {\bibfnamefont {J.}~\bibnamefont
  {Hickmann}}, \bibinfo {author} {\bibfnamefont {E.}~\bibnamefont {Fonseca}},
  \bibinfo {author} {\bibfnamefont {W.}~\bibnamefont {Soares}},\ and\ \bibinfo
  {author} {\bibfnamefont {S.}~\bibnamefont {Ch{\'a}vez-Cerda}},\ }\href@noop
  {} {\bibfield  {journal} {\bibinfo  {journal} {Physical Review Letters}\
  }\textbf {\bibinfo {volume} {105}},\ \bibinfo {pages} {053904} (\bibinfo
  {year} {2010})}\BibitemShut {NoStop}%
\bibitem [{\citenamefont {Dai}\ \emph {et~al.}(2015)\citenamefont {Dai},
  \citenamefont {Gao}, \citenamefont {Zhong}, \citenamefont {Na},\ and\
  \citenamefont {Wang}}]{dai2015measuring}%
  \BibitemOpen
  \bibfield  {author} {\bibinfo {author} {\bibfnamefont {K.}~\bibnamefont
  {Dai}}, \bibinfo {author} {\bibfnamefont {C.}~\bibnamefont {Gao}}, \bibinfo
  {author} {\bibfnamefont {L.}~\bibnamefont {Zhong}}, \bibinfo {author}
  {\bibfnamefont {Q.}~\bibnamefont {Na}},\ and\ \bibinfo {author}
  {\bibfnamefont {Q.}~\bibnamefont {Wang}},\ }\href@noop {} {\bibfield
  {journal} {\bibinfo  {journal} {Optics Letters}\ }\textbf {\bibinfo {volume}
  {40}},\ \bibinfo {pages} {562} (\bibinfo {year} {2015})}\BibitemShut
  {NoStop}%
\bibitem [{\citenamefont {Zheng}\ and\ \citenamefont
  {Wang}(2017)}]{zheng2017measuring}%
  \BibitemOpen
  \bibfield  {author} {\bibinfo {author} {\bibfnamefont {S.}~\bibnamefont
  {Zheng}}\ and\ \bibinfo {author} {\bibfnamefont {J.}~\bibnamefont {Wang}},\
  }\href@noop {} {\bibfield  {journal} {\bibinfo  {journal} {Scientific
  Reports}\ }\textbf {\bibinfo {volume} {7}},\ \bibinfo {pages} {1} (\bibinfo
  {year} {2017})}\BibitemShut {NoStop}%
\bibitem [{\citenamefont {Leach}\ \emph {et~al.}(2002)\citenamefont {Leach},
  \citenamefont {Padgett}, \citenamefont {Barnett}, \citenamefont
  {Franke-Arnold},\ and\ \citenamefont {Courtial}}]{leach2002measuring}%
  \BibitemOpen
  \bibfield  {author} {\bibinfo {author} {\bibfnamefont {J.}~\bibnamefont
  {Leach}}, \bibinfo {author} {\bibfnamefont {M.~J.}\ \bibnamefont {Padgett}},
  \bibinfo {author} {\bibfnamefont {S.~M.}\ \bibnamefont {Barnett}}, \bibinfo
  {author} {\bibfnamefont {S.}~\bibnamefont {Franke-Arnold}},\ and\ \bibinfo
  {author} {\bibfnamefont {J.}~\bibnamefont {Courtial}},\ }\href@noop {}
  {\bibfield  {journal} {\bibinfo  {journal} {Physical Review Letters}\
  }\textbf {\bibinfo {volume} {88}},\ \bibinfo {pages} {257901} (\bibinfo
  {year} {2002})}\BibitemShut {NoStop}%
\bibitem [{\citenamefont {Berkhout}\ \emph {et~al.}(2010)\citenamefont
  {Berkhout}, \citenamefont {Lavery}, \citenamefont {Courtial}, \citenamefont
  {Beijersbergen},\ and\ \citenamefont {Padgett}}]{berkhout2010efficient}%
  \BibitemOpen
  \bibfield  {author} {\bibinfo {author} {\bibfnamefont {G.~C.}\ \bibnamefont
  {Berkhout}}, \bibinfo {author} {\bibfnamefont {M.~P.}\ \bibnamefont
  {Lavery}}, \bibinfo {author} {\bibfnamefont {J.}~\bibnamefont {Courtial}},
  \bibinfo {author} {\bibfnamefont {M.~W.}\ \bibnamefont {Beijersbergen}},\
  and\ \bibinfo {author} {\bibfnamefont {M.~J.}\ \bibnamefont {Padgett}},\
  }\href@noop {} {\bibfield  {journal} {\bibinfo  {journal} {Physical Review
  Letters}\ }\textbf {\bibinfo {volume} {105}},\ \bibinfo {pages} {153601}
  (\bibinfo {year} {2010})}\BibitemShut {NoStop}%
\bibitem [{\citenamefont {Lavery}\ \emph {et~al.}(2012)\citenamefont {Lavery},
  \citenamefont {Robertson}, \citenamefont {Berkhout}, \citenamefont {Love},
  \citenamefont {Padgett},\ and\ \citenamefont
  {Courtial}}]{lavery2012refractive}%
  \BibitemOpen
  \bibfield  {author} {\bibinfo {author} {\bibfnamefont {M.~P.}\ \bibnamefont
  {Lavery}}, \bibinfo {author} {\bibfnamefont {D.~J.}\ \bibnamefont
  {Robertson}}, \bibinfo {author} {\bibfnamefont {G.~C.}\ \bibnamefont
  {Berkhout}}, \bibinfo {author} {\bibfnamefont {G.~D.}\ \bibnamefont {Love}},
  \bibinfo {author} {\bibfnamefont {M.~J.}\ \bibnamefont {Padgett}},\ and\
  \bibinfo {author} {\bibfnamefont {J.}~\bibnamefont {Courtial}},\ }\href@noop
  {} {\bibfield  {journal} {\bibinfo  {journal} {Optics Express}\ }\textbf
  {\bibinfo {volume} {20}},\ \bibinfo {pages} {2110} (\bibinfo {year}
  {2012})}\BibitemShut {NoStop}%
\bibitem [{\citenamefont {Mirhosseini}\ \emph {et~al.}(2013)\citenamefont
  {Mirhosseini}, \citenamefont {Malik}, \citenamefont {Shi},\ and\
  \citenamefont {Boyd}}]{mirhosseini2013efficient}%
  \BibitemOpen
  \bibfield  {author} {\bibinfo {author} {\bibfnamefont {M.}~\bibnamefont
  {Mirhosseini}}, \bibinfo {author} {\bibfnamefont {M.}~\bibnamefont {Malik}},
  \bibinfo {author} {\bibfnamefont {Z.}~\bibnamefont {Shi}},\ and\ \bibinfo
  {author} {\bibfnamefont {R.~W.}\ \bibnamefont {Boyd}},\ }\href@noop {}
  {\bibfield  {journal} {\bibinfo  {journal} {Nature Communications}\ }\textbf
  {\bibinfo {volume} {4}},\ \bibinfo {pages} {1} (\bibinfo {year}
  {2013})}\BibitemShut {NoStop}%
\bibitem [{\citenamefont {Wen}\ \emph {et~al.}(2018)\citenamefont {Wen},
  \citenamefont {Chremmos}, \citenamefont {Chen}, \citenamefont {Zhu},
  \citenamefont {Zhang},\ and\ \citenamefont {Yu}}]{wen2018spiral}%
  \BibitemOpen
  \bibfield  {author} {\bibinfo {author} {\bibfnamefont {Y.}~\bibnamefont
  {Wen}}, \bibinfo {author} {\bibfnamefont {I.}~\bibnamefont {Chremmos}},
  \bibinfo {author} {\bibfnamefont {Y.}~\bibnamefont {Chen}}, \bibinfo {author}
  {\bibfnamefont {J.}~\bibnamefont {Zhu}}, \bibinfo {author} {\bibfnamefont
  {Y.}~\bibnamefont {Zhang}},\ and\ \bibinfo {author} {\bibfnamefont
  {S.}~\bibnamefont {Yu}},\ }\href@noop {} {\bibfield  {journal} {\bibinfo
  {journal} {Physical Review Letters}\ }\textbf {\bibinfo {volume} {120}},\
  \bibinfo {pages} {193904} (\bibinfo {year} {2018})}\BibitemShut {NoStop}%
\bibitem [{\citenamefont {Labroille}\ \emph {et~al.}(2014)\citenamefont
  {Labroille}, \citenamefont {Denolle}, \citenamefont {Jian}, \citenamefont
  {Genevaux}, \citenamefont {Treps},\ and\ \citenamefont
  {Morizur}}]{labroille2014efficient}%
  \BibitemOpen
  \bibfield  {author} {\bibinfo {author} {\bibfnamefont {G.}~\bibnamefont
  {Labroille}}, \bibinfo {author} {\bibfnamefont {B.}~\bibnamefont {Denolle}},
  \bibinfo {author} {\bibfnamefont {P.}~\bibnamefont {Jian}}, \bibinfo {author}
  {\bibfnamefont {P.}~\bibnamefont {Genevaux}}, \bibinfo {author}
  {\bibfnamefont {N.}~\bibnamefont {Treps}},\ and\ \bibinfo {author}
  {\bibfnamefont {J.-F.}\ \bibnamefont {Morizur}},\ }\href@noop {} {\bibfield
  {journal} {\bibinfo  {journal} {Optics Express}\ }\textbf {\bibinfo {volume}
  {22}},\ \bibinfo {pages} {15599} (\bibinfo {year} {2014})}\BibitemShut
  {NoStop}%
\bibitem [{\citenamefont {Fontaine}\ \emph {et~al.}(2019)\citenamefont
  {Fontaine}, \citenamefont {Ryf}, \citenamefont {Chen}, \citenamefont
  {Neilson}, \citenamefont {Kim},\ and\ \citenamefont
  {Carpenter}}]{fontaine2019laguerre}%
  \BibitemOpen
  \bibfield  {author} {\bibinfo {author} {\bibfnamefont {N.~K.}\ \bibnamefont
  {Fontaine}}, \bibinfo {author} {\bibfnamefont {R.}~\bibnamefont {Ryf}},
  \bibinfo {author} {\bibfnamefont {H.}~\bibnamefont {Chen}}, \bibinfo {author}
  {\bibfnamefont {D.~T.}\ \bibnamefont {Neilson}}, \bibinfo {author}
  {\bibfnamefont {K.}~\bibnamefont {Kim}},\ and\ \bibinfo {author}
  {\bibfnamefont {J.}~\bibnamefont {Carpenter}},\ }\href@noop {} {\bibfield
  {journal} {\bibinfo  {journal} {Nature Communications}\ }\textbf {\bibinfo
  {volume} {10}},\ \bibinfo {pages} {1} (\bibinfo {year} {2019})}\BibitemShut
  {NoStop}%
\bibitem [{\citenamefont {Zhang}\ \emph {et~al.}(2020)\citenamefont {Zhang},
  \citenamefont {Wen}, \citenamefont {Fardoost}, \citenamefont {Fan},
  \citenamefont {Fontaine}, \citenamefont {Chen}, \citenamefont {Likamwa},\
  and\ \citenamefont {Li}}]{zhang2020simultaneous}%
  \BibitemOpen
  \bibfield  {author} {\bibinfo {author} {\bibfnamefont {Y.}~\bibnamefont
  {Zhang}}, \bibinfo {author} {\bibfnamefont {H.}~\bibnamefont {Wen}}, \bibinfo
  {author} {\bibfnamefont {A.}~\bibnamefont {Fardoost}}, \bibinfo {author}
  {\bibfnamefont {S.}~\bibnamefont {Fan}}, \bibinfo {author} {\bibfnamefont
  {N.~K.}\ \bibnamefont {Fontaine}}, \bibinfo {author} {\bibfnamefont
  {H.}~\bibnamefont {Chen}}, \bibinfo {author} {\bibfnamefont {P.~L.}\
  \bibnamefont {Likamwa}},\ and\ \bibinfo {author} {\bibfnamefont
  {G.}~\bibnamefont {Li}},\ }\href@noop {} {\bibfield  {journal} {\bibinfo
  {journal} {arXiv preprint arXiv:2010.04859}\ } (\bibinfo {year}
  {2020})}\BibitemShut {NoStop}%
\bibitem [{\citenamefont {Huang}\ \emph {et~al.}(2013)\citenamefont {Huang},
  \citenamefont {Ren}, \citenamefont {Yan}, \citenamefont {Ahmed},
  \citenamefont {Yue}, \citenamefont {Bozovich}, \citenamefont {Erkmen},
  \citenamefont {Birnbaum}, \citenamefont {Dolinar}, \citenamefont {Tur} \emph
  {et~al.}}]{huang2013phase}%
  \BibitemOpen
  \bibfield  {author} {\bibinfo {author} {\bibfnamefont {H.}~\bibnamefont
  {Huang}}, \bibinfo {author} {\bibfnamefont {Y.}~\bibnamefont {Ren}}, \bibinfo
  {author} {\bibfnamefont {Y.}~\bibnamefont {Yan}}, \bibinfo {author}
  {\bibfnamefont {N.}~\bibnamefont {Ahmed}}, \bibinfo {author} {\bibfnamefont
  {Y.}~\bibnamefont {Yue}}, \bibinfo {author} {\bibfnamefont {A.}~\bibnamefont
  {Bozovich}}, \bibinfo {author} {\bibfnamefont {B.~I.}\ \bibnamefont
  {Erkmen}}, \bibinfo {author} {\bibfnamefont {K.}~\bibnamefont {Birnbaum}},
  \bibinfo {author} {\bibfnamefont {S.}~\bibnamefont {Dolinar}}, \bibinfo
  {author} {\bibfnamefont {M.}~\bibnamefont {Tur}}, \emph {et~al.},\
  }\href@noop {} {\bibfield  {journal} {\bibinfo  {journal} {Optics Letters}\
  }\textbf {\bibinfo {volume} {38}},\ \bibinfo {pages} {2348} (\bibinfo {year}
  {2013})}\BibitemShut {NoStop}%
\bibitem [{\citenamefont {Zhou}\ \emph {et~al.}(2017)\citenamefont {Zhou},
  \citenamefont {Fu}, \citenamefont {Dong}, \citenamefont {Zhang},
  \citenamefont {Chen}, \citenamefont {Cai}, \citenamefont {Li},\ and\
  \citenamefont {Zhang}}]{zhou2017orbital}%
  \BibitemOpen
  \bibfield  {author} {\bibinfo {author} {\bibfnamefont {H.-L.}\ \bibnamefont
  {Zhou}}, \bibinfo {author} {\bibfnamefont {D.-Z.}\ \bibnamefont {Fu}},
  \bibinfo {author} {\bibfnamefont {J.-J.}\ \bibnamefont {Dong}}, \bibinfo
  {author} {\bibfnamefont {P.}~\bibnamefont {Zhang}}, \bibinfo {author}
  {\bibfnamefont {D.-X.}\ \bibnamefont {Chen}}, \bibinfo {author}
  {\bibfnamefont {X.-L.}\ \bibnamefont {Cai}}, \bibinfo {author} {\bibfnamefont
  {F.-L.}\ \bibnamefont {Li}},\ and\ \bibinfo {author} {\bibfnamefont {X.-L.}\
  \bibnamefont {Zhang}},\ }\href@noop {} {\bibfield  {journal} {\bibinfo
  {journal} {Light: Science \& Applications}\ }\textbf {\bibinfo {volume}
  {6}},\ \bibinfo {pages} {e16251} (\bibinfo {year} {2017})}\BibitemShut
  {NoStop}%
\bibitem [{\citenamefont {D’Errico}\ \emph {et~al.}(2017)\citenamefont
  {D’Errico}, \citenamefont {D’Amelio}, \citenamefont {Piccirillo},
  \citenamefont {Cardano},\ and\ \citenamefont {Marrucci}}]{d2017measuring}%
  \BibitemOpen
  \bibfield  {author} {\bibinfo {author} {\bibfnamefont {A.}~\bibnamefont
  {D’Errico}}, \bibinfo {author} {\bibfnamefont {R.}~\bibnamefont
  {D’Amelio}}, \bibinfo {author} {\bibfnamefont {B.}~\bibnamefont
  {Piccirillo}}, \bibinfo {author} {\bibfnamefont {F.}~\bibnamefont
  {Cardano}},\ and\ \bibinfo {author} {\bibfnamefont {L.}~\bibnamefont
  {Marrucci}},\ }\href@noop {} {\bibfield  {journal} {\bibinfo  {journal}
  {Optica}\ }\textbf {\bibinfo {volume} {4}},\ \bibinfo {pages} {1350}
  (\bibinfo {year} {2017})}\BibitemShut {NoStop}%
\bibitem [{\citenamefont {Fu}\ \emph {et~al.}(2020)\citenamefont {Fu},
  \citenamefont {Zhai}, \citenamefont {Zhang}, \citenamefont {Liu},
  \citenamefont {Song}, \citenamefont {Zhou},\ and\ \citenamefont
  {Gao}}]{fu2020universal}%
  \BibitemOpen
  \bibfield  {author} {\bibinfo {author} {\bibfnamefont {S.}~\bibnamefont
  {Fu}}, \bibinfo {author} {\bibfnamefont {Y.}~\bibnamefont {Zhai}}, \bibinfo
  {author} {\bibfnamefont {J.}~\bibnamefont {Zhang}}, \bibinfo {author}
  {\bibfnamefont {X.}~\bibnamefont {Liu}}, \bibinfo {author} {\bibfnamefont
  {R.}~\bibnamefont {Song}}, \bibinfo {author} {\bibfnamefont {H.}~\bibnamefont
  {Zhou}},\ and\ \bibinfo {author} {\bibfnamefont {C.}~\bibnamefont {Gao}},\
  }\href@noop {} {\bibfield  {journal} {\bibinfo  {journal} {PhotoniX}\
  }\textbf {\bibinfo {volume} {1}},\ \bibinfo {pages} {1} (\bibinfo {year}
  {2020})}\BibitemShut {NoStop}%
\bibitem [{\citenamefont {Kulkarni}\ \emph {et~al.}(2017)\citenamefont
  {Kulkarni}, \citenamefont {Sahu}, \citenamefont {Maga{\~n}a-Loaiza},
  \citenamefont {Boyd},\ and\ \citenamefont {Jha}}]{kulkarni2017single}%
  \BibitemOpen
  \bibfield  {author} {\bibinfo {author} {\bibfnamefont {G.}~\bibnamefont
  {Kulkarni}}, \bibinfo {author} {\bibfnamefont {R.}~\bibnamefont {Sahu}},
  \bibinfo {author} {\bibfnamefont {O.~S.}\ \bibnamefont {Maga{\~n}a-Loaiza}},
  \bibinfo {author} {\bibfnamefont {R.~W.}\ \bibnamefont {Boyd}},\ and\
  \bibinfo {author} {\bibfnamefont {A.~K.}\ \bibnamefont {Jha}},\ }\href@noop
  {} {\bibfield  {journal} {\bibinfo  {journal} {Nature Communications}\
  }\textbf {\bibinfo {volume} {8}},\ \bibinfo {pages} {1} (\bibinfo {year}
  {2017})}\BibitemShut {NoStop}%
\bibitem [{\citenamefont {Ip}\ \emph {et~al.}(2008)\citenamefont {Ip},
  \citenamefont {Lau}, \citenamefont {Barros},\ and\ \citenamefont
  {Kahn}}]{ip2008coherent}%
  \BibitemOpen
  \bibfield  {author} {\bibinfo {author} {\bibfnamefont {E.}~\bibnamefont
  {Ip}}, \bibinfo {author} {\bibfnamefont {A.~P.~T.}\ \bibnamefont {Lau}},
  \bibinfo {author} {\bibfnamefont {D.~J.}\ \bibnamefont {Barros}},\ and\
  \bibinfo {author} {\bibfnamefont {J.~M.}\ \bibnamefont {Kahn}},\ }\href@noop
  {} {\bibfield  {journal} {\bibinfo  {journal} {Optics Express}\ }\textbf
  {\bibinfo {volume} {16}},\ \bibinfo {pages} {753} (\bibinfo {year}
  {2008})}\BibitemShut {NoStop}%
\bibitem [{\citenamefont {Peng}\ \emph {et~al.}(2009)\citenamefont {Peng},
  \citenamefont {Wu}, \citenamefont {Feng}, \citenamefont {Arbab},
  \citenamefont {Shamee}, \citenamefont {Yang}, \citenamefont {Christen},
  \citenamefont {Willner},\ and\ \citenamefont {Chi}}]{peng2009spectrally}%
  \BibitemOpen
  \bibfield  {author} {\bibinfo {author} {\bibfnamefont {W.-R.}\ \bibnamefont
  {Peng}}, \bibinfo {author} {\bibfnamefont {X.}~\bibnamefont {Wu}}, \bibinfo
  {author} {\bibfnamefont {K.-M.}\ \bibnamefont {Feng}}, \bibinfo {author}
  {\bibfnamefont {V.~R.}\ \bibnamefont {Arbab}}, \bibinfo {author}
  {\bibfnamefont {B.}~\bibnamefont {Shamee}}, \bibinfo {author} {\bibfnamefont
  {J.-Y.}\ \bibnamefont {Yang}}, \bibinfo {author} {\bibfnamefont {L.~C.}\
  \bibnamefont {Christen}}, \bibinfo {author} {\bibfnamefont {A.~E.}\
  \bibnamefont {Willner}},\ and\ \bibinfo {author} {\bibfnamefont
  {S.}~\bibnamefont {Chi}},\ }\href@noop {} {\bibfield  {journal} {\bibinfo
  {journal} {Optics Express}\ }\textbf {\bibinfo {volume} {17}},\ \bibinfo
  {pages} {9099} (\bibinfo {year} {2009})}\BibitemShut {NoStop}%
\bibitem [{\citenamefont {Randel}\ \emph {et~al.}(2015)\citenamefont {Randel},
  \citenamefont {Pilori}, \citenamefont {Chandrasekhar}, \citenamefont
  {Raybon},\ and\ \citenamefont {Winzer}}]{randel2015100}%
  \BibitemOpen
  \bibfield  {author} {\bibinfo {author} {\bibfnamefont {S.}~\bibnamefont
  {Randel}}, \bibinfo {author} {\bibfnamefont {D.}~\bibnamefont {Pilori}},
  \bibinfo {author} {\bibfnamefont {S.}~\bibnamefont {Chandrasekhar}}, \bibinfo
  {author} {\bibfnamefont {G.}~\bibnamefont {Raybon}},\ and\ \bibinfo {author}
  {\bibfnamefont {P.}~\bibnamefont {Winzer}},\ }in\ \href@noop {} {\emph
  {\bibinfo {booktitle} {2015 European Conference on Optical Communication
  (ECOC)}}}\ (\bibinfo {organization} {IEEE},\ \bibinfo {year} {2015})\ pp.\
  \bibinfo {pages} {1--3}\BibitemShut {NoStop}%
\bibitem [{\citenamefont {Mecozzi}\ \emph {et~al.}(2016)\citenamefont
  {Mecozzi}, \citenamefont {Antonelli},\ and\ \citenamefont
  {Shtaif}}]{mecozzi2016kramers}%
  \BibitemOpen
  \bibfield  {author} {\bibinfo {author} {\bibfnamefont {A.}~\bibnamefont
  {Mecozzi}}, \bibinfo {author} {\bibfnamefont {C.}~\bibnamefont {Antonelli}},\
  and\ \bibinfo {author} {\bibfnamefont {M.}~\bibnamefont {Shtaif}},\
  }\href@noop {} {\bibfield  {journal} {\bibinfo  {journal} {Optica}\ }\textbf
  {\bibinfo {volume} {3}},\ \bibinfo {pages} {1220} (\bibinfo {year}
  {2016})}\BibitemShut {NoStop}%
\bibitem [{\citenamefont {Li}\ \emph {et~al.}(2017)\citenamefont {Li},
  \citenamefont {Erk{\i}l{\i}n{\c{c}}}, \citenamefont {Shi}, \citenamefont
  {Sillekens}, \citenamefont {Galdino}, \citenamefont {Thomsen}, \citenamefont
  {Bayvel},\ and\ \citenamefont {Killey}}]{li2017ssbi}%
  \BibitemOpen
  \bibfield  {author} {\bibinfo {author} {\bibfnamefont {Z.}~\bibnamefont
  {Li}}, \bibinfo {author} {\bibfnamefont {M.~S.}\ \bibnamefont
  {Erk{\i}l{\i}n{\c{c}}}}, \bibinfo {author} {\bibfnamefont {K.}~\bibnamefont
  {Shi}}, \bibinfo {author} {\bibfnamefont {E.}~\bibnamefont {Sillekens}},
  \bibinfo {author} {\bibfnamefont {L.}~\bibnamefont {Galdino}}, \bibinfo
  {author} {\bibfnamefont {B.~C.}\ \bibnamefont {Thomsen}}, \bibinfo {author}
  {\bibfnamefont {P.}~\bibnamefont {Bayvel}},\ and\ \bibinfo {author}
  {\bibfnamefont {R.~I.}\ \bibnamefont {Killey}},\ }\href@noop {} {\bibfield
  {journal} {\bibinfo  {journal} {Journal of Lightwave Technology}\ }\textbf
  {\bibinfo {volume} {35}},\ \bibinfo {pages} {1887} (\bibinfo {year}
  {2017})}\BibitemShut {NoStop}%
\bibitem [{\citenamefont {Bo}\ and\ \citenamefont {Kim}(2018)}]{bo2018kramers}%
  \BibitemOpen
  \bibfield  {author} {\bibinfo {author} {\bibfnamefont {T.}~\bibnamefont
  {Bo}}\ and\ \bibinfo {author} {\bibfnamefont {H.}~\bibnamefont {Kim}},\
  }\href@noop {} {\bibfield  {journal} {\bibinfo  {journal} {Optics Express}\
  }\textbf {\bibinfo {volume} {26}},\ \bibinfo {pages} {13810} (\bibinfo {year}
  {2018})}\BibitemShut {NoStop}%
\bibitem [{\citenamefont {Mecozzi}\ \emph {et~al.}(2019)\citenamefont
  {Mecozzi}, \citenamefont {Antonelli},\ and\ \citenamefont
  {Shtaif}}]{mecozzi2019kramers}%
  \BibitemOpen
  \bibfield  {author} {\bibinfo {author} {\bibfnamefont {A.}~\bibnamefont
  {Mecozzi}}, \bibinfo {author} {\bibfnamefont {C.}~\bibnamefont {Antonelli}},\
  and\ \bibinfo {author} {\bibfnamefont {M.}~\bibnamefont {Shtaif}},\
  }\href@noop {} {\bibfield  {journal} {\bibinfo  {journal} {Advances in Optics
  and Photonics}\ }\textbf {\bibinfo {volume} {11}},\ \bibinfo {pages} {480}
  (\bibinfo {year} {2019})}\BibitemShut {NoStop}%
\bibitem [{\citenamefont {Baek}\ \emph {et~al.}(2019)\citenamefont {Baek},
  \citenamefont {Lee}, \citenamefont {Shin},\ and\ \citenamefont
  {Park}}]{baek2019kramers}%
  \BibitemOpen
  \bibfield  {author} {\bibinfo {author} {\bibfnamefont {Y.}~\bibnamefont
  {Baek}}, \bibinfo {author} {\bibfnamefont {K.}~\bibnamefont {Lee}}, \bibinfo
  {author} {\bibfnamefont {S.}~\bibnamefont {Shin}},\ and\ \bibinfo {author}
  {\bibfnamefont {Y.}~\bibnamefont {Park}},\ }\href@noop {} {\bibfield
  {journal} {\bibinfo  {journal} {Optica}\ }\textbf {\bibinfo {volume} {6}},\
  \bibinfo {pages} {45} (\bibinfo {year} {2019})}\BibitemShut {NoStop}%
\bibitem [{\citenamefont {Xie}\ \emph {et~al.}(2017{\natexlab{a}})\citenamefont
  {Xie}, \citenamefont {Liu}, \citenamefont {Li}, \citenamefont {Ren},
  \citenamefont {Zhao}, \citenamefont {Yan}, \citenamefont {Ahmed},
  \citenamefont {Wang}, \citenamefont {Willner}, \citenamefont {Bao} \emph
  {et~al.}}]{xie2017spatial}%
  \BibitemOpen
  \bibfield  {author} {\bibinfo {author} {\bibfnamefont {G.}~\bibnamefont
  {Xie}}, \bibinfo {author} {\bibfnamefont {C.}~\bibnamefont {Liu}}, \bibinfo
  {author} {\bibfnamefont {L.}~\bibnamefont {Li}}, \bibinfo {author}
  {\bibfnamefont {Y.}~\bibnamefont {Ren}}, \bibinfo {author} {\bibfnamefont
  {Z.}~\bibnamefont {Zhao}}, \bibinfo {author} {\bibfnamefont {Y.}~\bibnamefont
  {Yan}}, \bibinfo {author} {\bibfnamefont {N.}~\bibnamefont {Ahmed}}, \bibinfo
  {author} {\bibfnamefont {Z.}~\bibnamefont {Wang}}, \bibinfo {author}
  {\bibfnamefont {A.~J.}\ \bibnamefont {Willner}}, \bibinfo {author}
  {\bibfnamefont {C.}~\bibnamefont {Bao}}, \emph {et~al.},\ }\href@noop {}
  {\bibfield  {journal} {\bibinfo  {journal} {Optics Letters}\ }\textbf
  {\bibinfo {volume} {42}},\ \bibinfo {pages} {991} (\bibinfo {year}
  {2017}{\natexlab{a}})}\BibitemShut {NoStop}%
\bibitem [{\citenamefont {Malik}\ \emph {et~al.}(2014)\citenamefont {Malik},
  \citenamefont {Mirhosseini}, \citenamefont {Lavery}, \citenamefont {Leach},
  \citenamefont {Padgett},\ and\ \citenamefont {Boyd}}]{malik2014direct}%
  \BibitemOpen
  \bibfield  {author} {\bibinfo {author} {\bibfnamefont {M.}~\bibnamefont
  {Malik}}, \bibinfo {author} {\bibfnamefont {M.}~\bibnamefont {Mirhosseini}},
  \bibinfo {author} {\bibfnamefont {M.~P.}\ \bibnamefont {Lavery}}, \bibinfo
  {author} {\bibfnamefont {J.}~\bibnamefont {Leach}}, \bibinfo {author}
  {\bibfnamefont {M.~J.}\ \bibnamefont {Padgett}},\ and\ \bibinfo {author}
  {\bibfnamefont {R.~W.}\ \bibnamefont {Boyd}},\ }\href@noop {} {\bibfield
  {journal} {\bibinfo  {journal} {Nature Communications}\ }\textbf {\bibinfo
  {volume} {5}},\ \bibinfo {pages} {1} (\bibinfo {year} {2014})}\BibitemShut
  {NoStop}%
\bibitem [{\citenamefont {Hu}\ \emph {et~al.}(2018)\citenamefont {Hu},
  \citenamefont {Br{\`e}s},\ and\ \citenamefont {Huang}}]{hu2018talbot}%
  \BibitemOpen
  \bibfield  {author} {\bibinfo {author} {\bibfnamefont {J.}~\bibnamefont
  {Hu}}, \bibinfo {author} {\bibfnamefont {C.-S.}\ \bibnamefont {Br{\`e}s}},\
  and\ \bibinfo {author} {\bibfnamefont {C.-B.}\ \bibnamefont {Huang}},\
  }\href@noop {} {\bibfield  {journal} {\bibinfo  {journal} {Optics Letters}\
  }\textbf {\bibinfo {volume} {43}},\ \bibinfo {pages} {4033} (\bibinfo {year}
  {2018})}\BibitemShut {NoStop}%
\bibitem [{\citenamefont {Lin}\ \emph {et~al.}(2021)\citenamefont {Lin},
  \citenamefont {Hu}, \citenamefont {Chen}, \citenamefont {Yu},\ and\
  \citenamefont {Br{\`e}s}}]{lin2021spectral}%
  \BibitemOpen
  \bibfield  {author} {\bibinfo {author} {\bibfnamefont {Z.}~\bibnamefont
  {Lin}}, \bibinfo {author} {\bibfnamefont {J.}~\bibnamefont {Hu}}, \bibinfo
  {author} {\bibfnamefont {Y.}~\bibnamefont {Chen}}, \bibinfo {author}
  {\bibfnamefont {S.}~\bibnamefont {Yu}},\ and\ \bibinfo {author}
  {\bibfnamefont {C.-S.}\ \bibnamefont {Br{\`e}s}},\ }\href@noop {} {\bibfield
  {journal} {\bibinfo  {journal} {APL Photonics}\ }\textbf {\bibinfo {volume}
  {6}},\ \bibinfo {pages} {111302} (\bibinfo {year} {2021})}\BibitemShut
  {NoStop}%
\bibitem [{\citenamefont {G{\"o}tte}\ \emph {et~al.}(2008)\citenamefont
  {G{\"o}tte}, \citenamefont {O’Holleran}, \citenamefont {Preece},
  \citenamefont {Flossmann}, \citenamefont {Franke-Arnold}, \citenamefont
  {Barnett},\ and\ \citenamefont {Padgett}}]{gotte2008light}%
  \BibitemOpen
  \bibfield  {author} {\bibinfo {author} {\bibfnamefont {J.~B.}\ \bibnamefont
  {G{\"o}tte}}, \bibinfo {author} {\bibfnamefont {K.}~\bibnamefont
  {O’Holleran}}, \bibinfo {author} {\bibfnamefont {D.}~\bibnamefont
  {Preece}}, \bibinfo {author} {\bibfnamefont {F.}~\bibnamefont {Flossmann}},
  \bibinfo {author} {\bibfnamefont {S.}~\bibnamefont {Franke-Arnold}}, \bibinfo
  {author} {\bibfnamefont {S.~M.}\ \bibnamefont {Barnett}},\ and\ \bibinfo
  {author} {\bibfnamefont {M.~J.}\ \bibnamefont {Padgett}},\ }\href@noop {}
  {\bibfield  {journal} {\bibinfo  {journal} {Optics Express}\ }\textbf
  {\bibinfo {volume} {16}},\ \bibinfo {pages} {993} (\bibinfo {year}
  {2008})}\BibitemShut {NoStop}%
\bibitem [{\citenamefont {Vaity}\ and\ \citenamefont
  {Rusch}(2015)}]{vaity2015perfect}%
  \BibitemOpen
  \bibfield  {author} {\bibinfo {author} {\bibfnamefont {P.}~\bibnamefont
  {Vaity}}\ and\ \bibinfo {author} {\bibfnamefont {L.}~\bibnamefont {Rusch}},\
  }\href@noop {} {\bibfield  {journal} {\bibinfo  {journal} {Optics Letters}\
  }\textbf {\bibinfo {volume} {40}},\ \bibinfo {pages} {597} (\bibinfo {year}
  {2015})}\BibitemShut {NoStop}%
\bibitem [{\citenamefont {Mecozzi}(2016)}]{mecozzi2016necessary}%
  \BibitemOpen
  \bibfield  {author} {\bibinfo {author} {\bibfnamefont {A.}~\bibnamefont
  {Mecozzi}},\ }\href@noop {} {\bibfield  {journal} {\bibinfo  {journal} {arXiv
  preprint arXiv:1606.04861}\ } (\bibinfo {year} {2016})}\BibitemShut {NoStop}%
\bibitem [{\citenamefont {Cizek}(1970)}]{cizek1970discrete}%
  \BibitemOpen
  \bibfield  {author} {\bibinfo {author} {\bibfnamefont {V.}~\bibnamefont
  {Cizek}},\ }\href@noop {} {\bibfield  {journal} {\bibinfo  {journal} {IEEE
  Transactions on Audio and Electroacoustics}\ }\textbf {\bibinfo {volume}
  {18}},\ \bibinfo {pages} {340} (\bibinfo {year} {1970})}\BibitemShut
  {NoStop}%
\bibitem [{\citenamefont {Arriz{\'o}n}\ \emph {et~al.}(2007)\citenamefont
  {Arriz{\'o}n}, \citenamefont {Ruiz}, \citenamefont {Carrada},\ and\
  \citenamefont {Gonz{\'a}lez}}]{arrizon2007pixelated}%
  \BibitemOpen
  \bibfield  {author} {\bibinfo {author} {\bibfnamefont {V.}~\bibnamefont
  {Arriz{\'o}n}}, \bibinfo {author} {\bibfnamefont {U.}~\bibnamefont {Ruiz}},
  \bibinfo {author} {\bibfnamefont {R.}~\bibnamefont {Carrada}},\ and\ \bibinfo
  {author} {\bibfnamefont {L.~A.}\ \bibnamefont {Gonz{\'a}lez}},\ }\href@noop
  {} {\bibfield  {journal} {\bibinfo  {journal} {JOSA A}\ }\textbf {\bibinfo
  {volume} {24}},\ \bibinfo {pages} {3500} (\bibinfo {year}
  {2007})}\BibitemShut {NoStop}%
\bibitem [{\citenamefont {Xie}\ \emph {et~al.}(2017{\natexlab{b}})\citenamefont
  {Xie}, \citenamefont {Song}, \citenamefont {Zhao}, \citenamefont {Milione},
  \citenamefont {Ren}, \citenamefont {Liu}, \citenamefont {Zhang},
  \citenamefont {Bao}, \citenamefont {Li}, \citenamefont {Wang} \emph
  {et~al.}}]{xie2017using}%
  \BibitemOpen
  \bibfield  {author} {\bibinfo {author} {\bibfnamefont {G.}~\bibnamefont
  {Xie}}, \bibinfo {author} {\bibfnamefont {H.}~\bibnamefont {Song}}, \bibinfo
  {author} {\bibfnamefont {Z.}~\bibnamefont {Zhao}}, \bibinfo {author}
  {\bibfnamefont {G.}~\bibnamefont {Milione}}, \bibinfo {author} {\bibfnamefont
  {Y.}~\bibnamefont {Ren}}, \bibinfo {author} {\bibfnamefont {C.}~\bibnamefont
  {Liu}}, \bibinfo {author} {\bibfnamefont {R.}~\bibnamefont {Zhang}}, \bibinfo
  {author} {\bibfnamefont {C.}~\bibnamefont {Bao}}, \bibinfo {author}
  {\bibfnamefont {L.}~\bibnamefont {Li}}, \bibinfo {author} {\bibfnamefont
  {Z.}~\bibnamefont {Wang}}, \emph {et~al.},\ }\href@noop {} {\bibfield
  {journal} {\bibinfo  {journal} {Optics Letters}\ }\textbf {\bibinfo {volume}
  {42}},\ \bibinfo {pages} {4482} (\bibinfo {year}
  {2017}{\natexlab{b}})}\BibitemShut {NoStop}%
\bibitem [{\citenamefont {De~Chatellus}\ \emph {et~al.}(2015)\citenamefont
  {De~Chatellus}, \citenamefont {Lacot}, \citenamefont {Hugon}, \citenamefont
  {Jacquin}, \citenamefont {Khebbache},\ and\ \citenamefont
  {Aza{\~n}a}}]{de2015phases}%
  \BibitemOpen
  \bibfield  {author} {\bibinfo {author} {\bibfnamefont {H.~G.}\ \bibnamefont
  {De~Chatellus}}, \bibinfo {author} {\bibfnamefont {E.}~\bibnamefont {Lacot}},
  \bibinfo {author} {\bibfnamefont {O.}~\bibnamefont {Hugon}}, \bibinfo
  {author} {\bibfnamefont {O.}~\bibnamefont {Jacquin}}, \bibinfo {author}
  {\bibfnamefont {N.}~\bibnamefont {Khebbache}},\ and\ \bibinfo {author}
  {\bibfnamefont {J.}~\bibnamefont {Aza{\~n}a}},\ }\href@noop {} {\bibfield
  {journal} {\bibinfo  {journal} {JOSA A}\ }\textbf {\bibinfo {volume} {32}},\
  \bibinfo {pages} {1132} (\bibinfo {year} {2015})}\BibitemShut {NoStop}%
\bibitem [{\citenamefont {Clement}\ \emph {et~al.}(2020)\citenamefont
  {Clement}, \citenamefont {de~Chatellus},\ and\ \citenamefont
  {Fern{\'a}ndez-Pousa}}]{clement2020far}%
  \BibitemOpen
  \bibfield  {author} {\bibinfo {author} {\bibfnamefont {J.}~\bibnamefont
  {Clement}}, \bibinfo {author} {\bibfnamefont {H.~G.}\ \bibnamefont
  {de~Chatellus}},\ and\ \bibinfo {author} {\bibfnamefont {C.~R.}\ \bibnamefont
  {Fern{\'a}ndez-Pousa}},\ }\href@noop {} {\bibfield  {journal} {\bibinfo
  {journal} {Optics Express}\ }\textbf {\bibinfo {volume} {28}},\ \bibinfo
  {pages} {12977} (\bibinfo {year} {2020})}\BibitemShut {NoStop}%
\bibitem [{\citenamefont {Zhang}\ \emph {et~al.}(2022)\citenamefont {Zhang},
  \citenamefont {Zeng}, \citenamefont {Lu}, \citenamefont {Wang}, \citenamefont
  {Zhao},\ and\ \citenamefont {Cai}}]{zhang2022review}%
  \BibitemOpen
  \bibfield  {author} {\bibinfo {author} {\bibfnamefont {H.}~\bibnamefont
  {Zhang}}, \bibinfo {author} {\bibfnamefont {J.}~\bibnamefont {Zeng}},
  \bibinfo {author} {\bibfnamefont {X.}~\bibnamefont {Lu}}, \bibinfo {author}
  {\bibfnamefont {Z.}~\bibnamefont {Wang}}, \bibinfo {author} {\bibfnamefont
  {C.}~\bibnamefont {Zhao}},\ and\ \bibinfo {author} {\bibfnamefont
  {Y.}~\bibnamefont {Cai}},\ }\href@noop {} {\bibfield  {journal} {\bibinfo
  {journal} {Nanophotonics}\ }\textbf {\bibinfo {volume} {11}},\ \bibinfo
  {pages} {241} (\bibinfo {year} {2022})}\BibitemShut {NoStop}%
\bibitem [{\citenamefont {Courtial}\ \emph {et~al.}(1998)\citenamefont
  {Courtial}, \citenamefont {Dholakia}, \citenamefont {Robertson},
  \citenamefont {Allen},\ and\ \citenamefont
  {Padgett}}]{courtial1998measurement}%
  \BibitemOpen
  \bibfield  {author} {\bibinfo {author} {\bibfnamefont {J.}~\bibnamefont
  {Courtial}}, \bibinfo {author} {\bibfnamefont {K.}~\bibnamefont {Dholakia}},
  \bibinfo {author} {\bibfnamefont {D.}~\bibnamefont {Robertson}}, \bibinfo
  {author} {\bibfnamefont {L.}~\bibnamefont {Allen}},\ and\ \bibinfo {author}
  {\bibfnamefont {M.}~\bibnamefont {Padgett}},\ }\href@noop {} {\bibfield
  {journal} {\bibinfo  {journal} {Physical Review Letters}\ }\textbf {\bibinfo
  {volume} {80}},\ \bibinfo {pages} {3217} (\bibinfo {year}
  {1998})}\BibitemShut {NoStop}%
\end{thebibliography}%


\begin{thebibliography}{4}%
\makeatletter
\providecommand \@ifxundefined [1]{%
 \@ifx{#1\undefined}
}%
\providecommand \@ifnum [1]{%
 \ifnum #1\expandafter \@firstoftwo
 \else \expandafter \@secondoftwo
 \fi
}%
\providecommand \@ifx [1]{%
 \ifx #1\expandafter \@firstoftwo
 \else \expandafter \@secondoftwo
 \fi
}%
\providecommand \natexlab [1]{#1}%
\providecommand \enquote  [1]{``#1''}%
\providecommand \bibnamefont  [1]{#1}%
\providecommand \bibfnamefont [1]{#1}%
\providecommand \citenamefont [1]{#1}%
\providecommand \href@noop [0]{\@secondoftwo}%
\providecommand \href [0]{\begingroup \@sanitize@url \@href}%
\providecommand \@href[1]{\@@startlink{#1}\@@href}%
\providecommand \@@href[1]{\endgroup#1\@@endlink}%
\providecommand \@sanitize@url [0]{\catcode `\\12\catcode `\$12\catcode
  `\&12\catcode `\#12\catcode `\^12\catcode `\_12\catcode `\%12\relax}%
\providecommand \@@startlink[1]{}%
\providecommand \@@endlink[0]{}%
\providecommand \url  [0]{\begingroup\@sanitize@url \@url }%
\providecommand \@url [1]{\endgroup\@href {#1}{\urlprefix }}%
\providecommand \urlprefix  [0]{URL }%
\providecommand \Eprint [0]{\href }%
\providecommand \doibase [0]{https://doi.org/}%
\providecommand \selectlanguage [0]{\@gobble}%
\providecommand \bibinfo  [0]{\@secondoftwo}%
\providecommand \bibfield  [0]{\@secondoftwo}%
\providecommand \translation [1]{[#1]}%
\providecommand \BibitemOpen [0]{}%
\providecommand \bibitemStop [0]{}%
\providecommand \bibitemNoStop [0]{.\EOS\space}%
\providecommand \EOS [0]{\spacefactor3000\relax}%
\providecommand \BibitemShut  [1]{\csname bibitem#1\endcsname}%
\let\auto@bib@innerbib\@empty
%</preamble>
\bibitem [{\citenamefont {Arriz{\'o}n}\ \emph {et~al.}(2007)\citenamefont
  {Arriz{\'o}n}, \citenamefont {Ruiz}, \citenamefont {Carrada},\ and\
  \citenamefont {Gonz{\'a}lez}}]{arrizon2007pixelated}%
  \BibitemOpen
  \bibfield  {author} {\bibinfo {author} {\bibfnamefont {V.}~\bibnamefont
  {Arriz{\'o}n}}, \bibinfo {author} {\bibfnamefont {U.}~\bibnamefont {Ruiz}},
  \bibinfo {author} {\bibfnamefont {R.}~\bibnamefont {Carrada}},\ and\ \bibinfo
  {author} {\bibfnamefont {L.~A.}\ \bibnamefont {Gonz{\'a}lez}},\ }\href@noop
  {} {\bibfield  {journal} {\bibinfo  {journal} {JOSA A}\ }\textbf {\bibinfo
  {volume} {24}},\ \bibinfo {pages} {3500} (\bibinfo {year}
  {2007})}\BibitemShut {NoStop}%
\bibitem [{\citenamefont {Vaity}\ and\ \citenamefont
  {Rusch}(2015)}]{vaity2015perfect}%
  \BibitemOpen
  \bibfield  {author} {\bibinfo {author} {\bibfnamefont {P.}~\bibnamefont
  {Vaity}}\ and\ \bibinfo {author} {\bibfnamefont {L.}~\bibnamefont {Rusch}},\
  }\href@noop {} {\bibfield  {journal} {\bibinfo  {journal} {Optics Letters}\
  }\textbf {\bibinfo {volume} {40}},\ \bibinfo {pages} {597} (\bibinfo {year}
  {2015})}\BibitemShut {NoStop}%
\bibitem [{\citenamefont {Mecozzi}\ \emph {et~al.}(2019)\citenamefont
  {Mecozzi}, \citenamefont {Antonelli},\ and\ \citenamefont
  {Shtaif}}]{mecozzi2019kramers}%
  \BibitemOpen
  \bibfield  {author} {\bibinfo {author} {\bibfnamefont {A.}~\bibnamefont
  {Mecozzi}}, \bibinfo {author} {\bibfnamefont {C.}~\bibnamefont {Antonelli}},\
  and\ \bibinfo {author} {\bibfnamefont {M.}~\bibnamefont {Shtaif}},\
  }\href@noop {} {\bibfield  {journal} {\bibinfo  {journal} {Advances in Optics
  and Photonics}\ }\textbf {\bibinfo {volume} {11}},\ \bibinfo {pages} {480}
  (\bibinfo {year} {2019})}\BibitemShut {NoStop}%
\bibitem [{\citenamefont {Baek}\ \emph {et~al.}(2019)\citenamefont {Baek},
  \citenamefont {Lee}, \citenamefont {Shin},\ and\ \citenamefont
  {Park}}]{baek2019kramers}%
  \BibitemOpen
  \bibfield  {author} {\bibinfo {author} {\bibfnamefont {Y.}~\bibnamefont
  {Baek}}, \bibinfo {author} {\bibfnamefont {K.}~\bibnamefont {Lee}}, \bibinfo
  {author} {\bibfnamefont {S.}~\bibnamefont {Shin}},\ and\ \bibinfo {author}
  {\bibfnamefont {Y.}~\bibnamefont {Park}},\ }\href@noop {} {\bibfield
  {journal} {\bibinfo  {journal} {Optica}\ }\textbf {\bibinfo {volume} {6}},\
  \bibinfo {pages} {45} (\bibinfo {year} {2019})}\BibitemShut {NoStop}%
\end{thebibliography}%

\end{document}

\title{Supplementary information for:\\Single-shot Kramers–Kronig complex orbital angular momentum spectrum retrieval}

\author{Zhongzheng Lin$^{1,\ast}$, 
        Jianqi Hu$^{2,\ast,\dagger}$, 
        Yujie Chen$^1$, 
        Camille-Sophie Br\`es$^2$, 
        and Siyuan Yu$^{1,\ddag}$}
\affiliation{
$^1$State Key Laboratory of Optoelectronic Materials and Technologies, School of Electronics and Information Technology, Sun Yat-sen University, Guangzhou 510275, China.\\
$^2${\'E}cole Polytechnique F{\'e}d{\'e}rale de Lausanne, Photonic Systems Laboratory (PHOSL), STI-IEM, Station 11, Lausanne CH-1015, Switzerland.}

\maketitle

%%%%%%%%%%%%%%%%%%%%%
\section*{\textbf{Supplementary Note 1. Experimental setup}}

\begin{figure}[htbp]
\renewcommand{\figurename}{Supplementary Figure}
\centering
\includegraphics[width=0.75\linewidth]{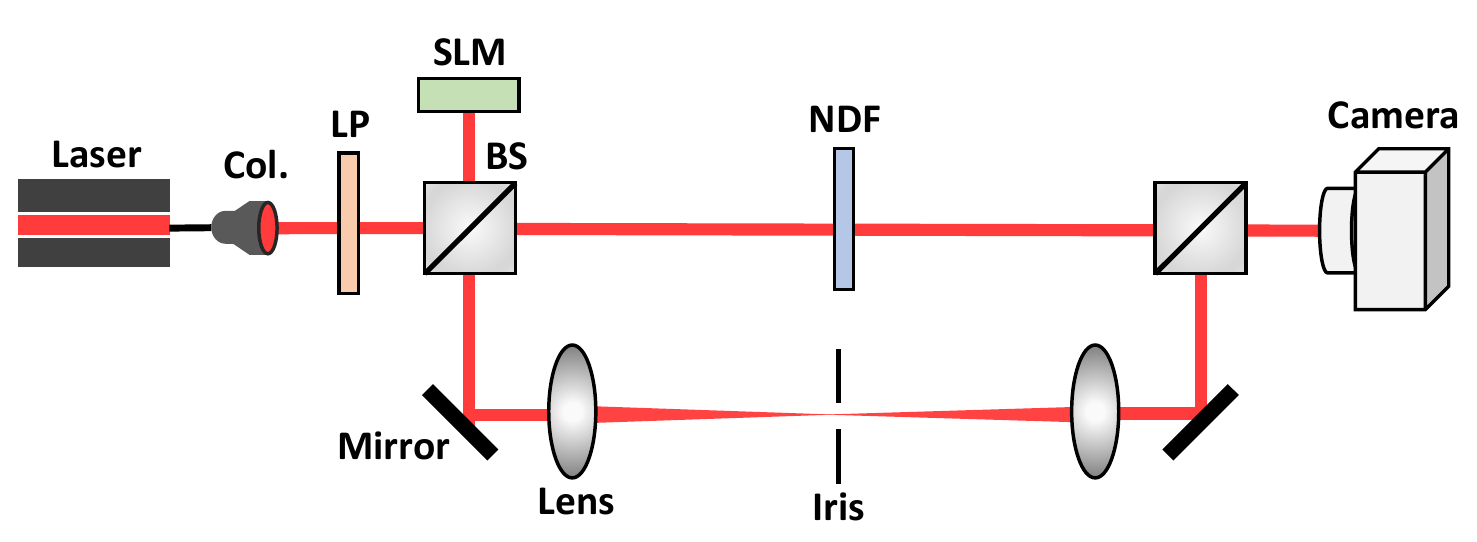}
\caption{Experimental setup for KK complex OAM spectrum retrieval. Col.: collimator; LP: linear polarizer; BS: beam splitter; SLM: spatial light modulator; NDF: neutral density filter.}
\label{fig:s1}
\end{figure}

Supplementary Figure \ref{fig:s1} illustrates the experimental setup used for Kramers-Kronig (KK) complex orbital angular momentum (OAM) spectrum retrieval. A continuous-wave laser at $1550~{\rm nm}$ is collimated, and polarization managed to match the working-axis of the spatial light modulator (SLM). The collimated beam is then split into the reference path and the signal path via a beam splitter. The signal field used in the experiment is prepared in a 4-f system with a SLM at its input plane and an iris at its Fourier plane. Such a configuration allows for the synthesis of arbitrary complex OAM fields \cite{arrizon2007pixelated}. The OAM field under test is co-axially combined with the reference beam via another beam splitter. A neutral density filter is inserted in the reference path to adjust its optical power. At the output of the $4$-f system, the camera captures the interferogram of the signal and the reference beams. 

The OAM mode basis used in this work is the ring-shaped perfect vortex mode, which can be approximately represented as \cite{vaity2015perfect}:
\begin{equation}
E(r,\phi)=e^{-\frac{(r-r_0)^2}{w_0^2}}e^{il\phi}
\label{eq:refname1}
\end{equation}
where $r_0$, $w_0$ are the radius and half-width of the ring, respectively. In the experiment, $r_0=750~\mu{\rm m}$, $w_0=150~\mu{\rm m}$ are chosen for the OAM measurement space with topological charges spanning from $1$ to $20$. When the dimensionality is enlarged to $30$, the beam parameters are readjusted to $r_0=1000~\mu{\rm m}$, $w_0=200~\mu{\rm m}$ .

\section*{\textbf{Supplementary Note 2. Control of the carrier-to-signal power ratio}}

In this section, we discuss how we experimentally control the carrier-to-signal power ratio (CSPR) in the OAM spectrum retrieval process. Noticeably, for different complex states, the CSPRs required to satisfy the minimum phase condition are different. In the following, we show that, by using the arbitrary OAM field synthesis method described above \cite{arrizon2007pixelated}, the difference between the experimental CSPR and the minimum required CSPR is well maintained automatically for all cases. 

The signal field is generated by carving the input Gaussian beam into the target field structure through a computer-generated hologram (CGH). Since the synthesized OAM states possess identical radial distributions, we only consider their distributions in the azimuthal angle: 
\begin{equation}
E_s (\phi)=\sqrt{P_0 \gamma} a(\phi)e^{i\varphi(\phi)},
\label{eq:refname2}
\end{equation}
where $P_0$ is the power of the ring-shaped beam with flat phase profile, and $\gamma$ denotes the diffraction efficiency that generates the CGH. $a(\phi)$ and $\varphi(\phi)$ represent the relative amplitude and phase distributions of the target OAM states, respectively. Here the amplitude $a(\phi)$ is normalized such that $\operatorname{max}\{a(\phi)\}=1$ for $\phi \in [0, 2\pi)$. To meet the minimum phase condition, the reference amplitude needs to be greater than the peak amplitude of the signal field [3]. This sets the minimum CSPR required for the measurement:
\begin{equation}
{\rm CSPR_m}= 10 \log \frac{\operatorname{max}{|E_s(\phi)|^2}}{P_s} =10\log \frac{P_c \gamma}{P_s},
\label{eq:refname3}
\end{equation}
where $P_s=<|E_s(\phi)|^2>$ is the power of the signal field average in the azimuthal angle. The difference between the experimental CSPR and the minimum required CSPR writes:
\begin{equation}
{\rm CSPR}-{\rm CSPR_m}= 10\log \frac{P_c}{P_s} - 10\log \frac{P_0 \gamma}{P_s} = 10\log \frac{P_c}{P_0 \gamma}.
\label{eq:refname4}
\end{equation}

\begin{figure}[htbp]
\renewcommand{\figurename}{Supplementary Figure}
\centering
\includegraphics[width=0.9\linewidth]{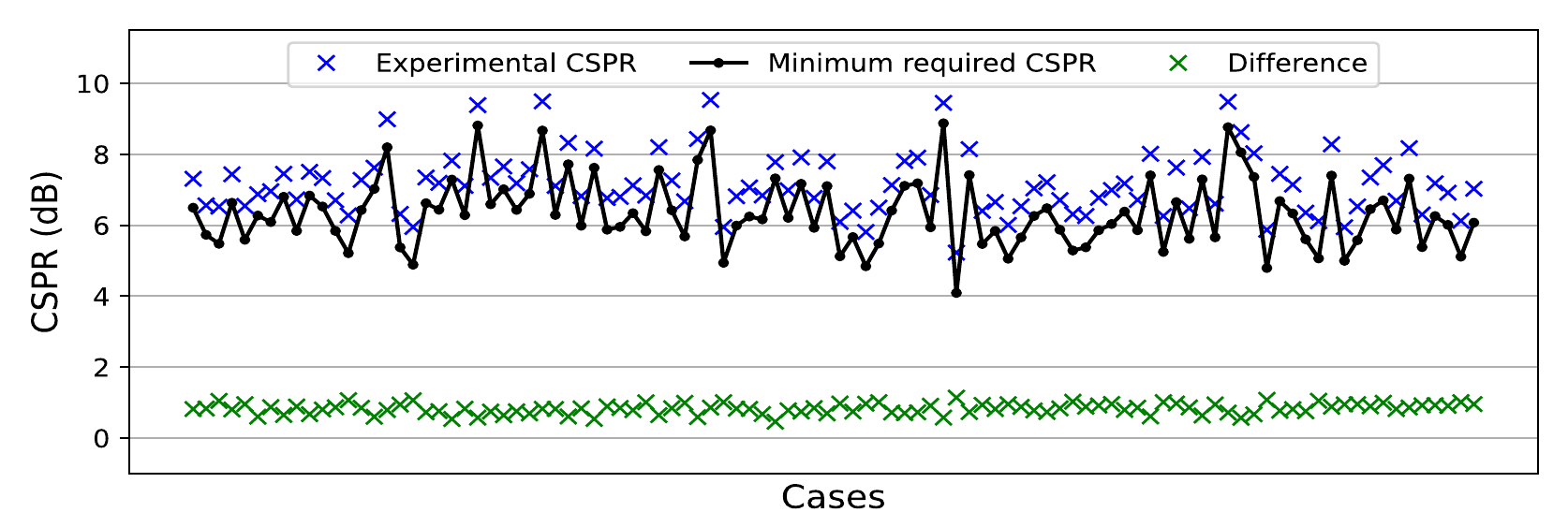}
\caption{Experimentally measured CSPRs, minimum required CSPRs and their differences for $100$ random complex OAM states. The experimental CSPRs are slightly higher than the minimum required CSPR values, and their differences are roughly kept the same for different cases.}
\label{fig:s2}
\end{figure}

We can see that the difference here only depends on the diffraction efficiency $\gamma$. By keeping $\gamma$ constant throughout the experiment, we can always ensure the experimental CSPR to be slightly higher than the minimum required CSPR value. In practice, we usually keep their difference around 1 dB. Supplementary Figure \ref{fig:s2} shows the CSPR conditions for $100$ random OAM fields [one tenth of the fields in Fig. 5 (a) in the main text]. The CSPR difference is well maintained for distinct OAM spectra.   

In scenarios where the experimental CSPR needs to be changed, instead of varying the attenuation for the reference beam, we can also adjust the diffraction efficiency $\gamma$ of the CGH. In this way, the signal power is varied thereby effectively changing the CSPR. This digital method allows for a more controlled adjustment of the experimental CSPR, which does not affect the alignment of the setup. For instance, when studying the retrieval performance at different CSPR levels (corresponding to Fig. 4 in the main text), the signal power is decreased in steps of $0.5~{\rm dB}$ by varying the diffraction efficiency $\gamma$.

\section*{\textbf{Supplementary Note 3. Effect of upsampling}}
The logarithmic operation taken in the KK retrieval procedure expands the bandwidth of the OAM spectrum \cite{mecozzi2019kramers}. If the number of physical sampling points in the azimuthal angle does not cover the broadened spectrum, digital upsampling is necessary to ensure accurate retrieval. The upsampling is implemented on the normalized interferogram $|E_i(\phi)|^2/|E_s(\phi)|^2$, by means of zero-padding in the Fourier domain \cite{baek2019kramers}. After the Hilbert transformation and necessary computations, the data is downsampled to the original number of samples. 

\begin{figure}[htp]
\renewcommand{\figurename}{Supplementary Figure}
\centering
\includegraphics[width=0.9\linewidth]{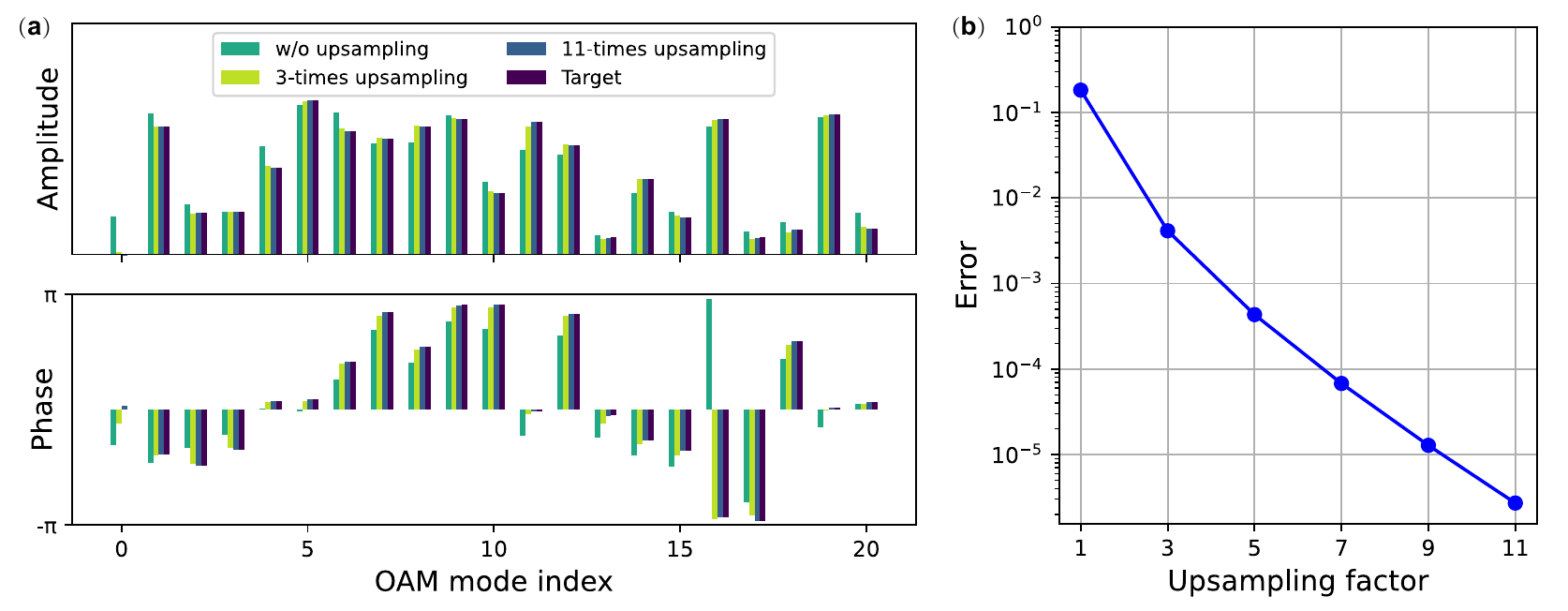}
\caption{Effect of digital upsampling on the KK retrieval performance. (a) The amplitude and phase of the retrieved OAM spectrum without, and with $3$-times and $11$-times digital upsampling. (b) The error of the retrieved complex OAM spectrum versus the upsampling factor. The retrieval accuracy improves with the increase of the digital upsampling.}
\label{fig:s3}
\end{figure}

In the following, we show the effect of upsampling in simulation, for retrieving the same complex OAM spectrum as in Fig. 2 of the main text. To demonstrate its full retrieval capability, the complex field with OAM mode indices from $1$ to $20$ is sampled at the Nyquist frequency (41 azimuthal samples). Supplementary Figure \ref{fig:s3}(a) shows the retrieved OAM spectra without and with upsampling of different factors. It can be seen that, without upsampling, the retrieved field cannot reproduce perfectly the target spectrum. The accuracy of the retrieval (defined as the overlap integral of the retrieved field and the ground truth) remains low in this case ($\sim 0.81$), as quantified in Supplementary Figure \ref{fig:s3}(b) with ${\rm error} = 1 - {\rm accuracy}$. However, only $3$-times digital upsampling could significantly boost the accuracy to $0.99$, which is also manifested by the retrieved OAM spectrum. While almost perfect retrieval can be reached by further increasing the upsampling factor to $11$-times. As shown in Supplementary Figure \ref{fig:s3}(b), the error is minimized and the accuracy is converged to unity with a higher upsampling factor. It is worth to mention that, unlike time signals in optical communications, the number of azimuthal samples here is a relatively small value given by the OAM measurement spectral range. As such, the digital upsampling does not add much computational complexity to the retrieval process. 

\renewcommand{\bibpreamble}{
$^\ast$These authors contributed equally to this work.\\
$^\dagger${Corresponding author: \textcolor{magenta}{jianqi.hu@epfl.ch}}\\
$^\ddag${Email: \textcolor{magenta}{yusy@mail.sysu.edu.cn}}
}
\pretolerance=0
\bigskip
\bibliographystyle{apsrev4-2}
\bibliography{ref}